\documentclass[twocolumn,prc,floatfix,showpacs,nopreprintnumbers,nofootinbib,%
superscriptaddress]{revtex4}
\usepackage{mathrsfs}
\usepackage{amssymb}
\usepackage{amsmath}
\usepackage{subdepth}
\usepackage{graphicx}
\usepackage{xcolor}
\definecolor{lcolor}{rgb}{0.5,0,0}
\definecolor{citcolor}{rgb}{0,0.3,0.0}

\usepackage[breaklinks,colorlinks,urlcolor=blue,citecolor=citcolor,linkcolor=lcolor]{hyperref}
\usepackage{mciteplus}
\usepackage[latin1]{inputenc}

\newcommand{\rt}{{\mathbf{r}}}
\newcommand{\xt}{{\mathbf{x}}}
\newcommand{\yt}{{\mathbf{y}}}
\newcommand{\ut}{{\mathbf{u}}}
\newcommand{\vt}{{\mathbf{v}}}
\newcommand{\zt}{{\mathbf{z}}}
\newcommand{\ztp}{{\mathbf{z}'}}
\newcommand{\xti}{{\mathbf{x}_{1}}}
\newcommand{\xtn}{{\mathbf{x}_{n}}}
\newcommand{\yti}{{\mathbf{y}_{1}}}
\newcommand{\ytn}{{\mathbf{y}_{n}}}
\newcommand{\qt}{{\mathbf{q}}}
\newcommand{\kt}{{\mathbf{k}}}
\newcommand{\pt}{{\mathbf{p}}}

\newcommand{\Kt}{{\mathbf{K}}}
\newcommand{\Pt}{{\mathbf{P}}}
\newcommand{\xit}{{\boldsymbol{\xi}}}
\newcommand{\etat}{{\boldsymbol{\eta}}}

\newcommand{\xitilt}{{\boldsymbol{\widetilde{\xi}}}}
\newcommand{\astil}{{ \widetilde{\alpha} }}

\newcommand{\ptt}{p_T} 
\newcommand{\ktt}{k_T} 
\newcommand{\qtt}{q_T} 
\newcommand{\xtt}{x_T} 
\newcommand{\rtt}{r_T} 

\newcommand{\ud}{\, \mathrm{d}}

\newcommand{\tr}{\, \mathrm{Tr} \, }

\newcommand{\nc}{{N_\mathrm{c}}}
\newcommand{\nf}{{N_\mathrm{F}}}

\newcommand{\nr}[1]{(\ref{#1})}

\newcommand{\gev}{\ \textrm{GeV}}

\newcommand{\qs}{Q_\mathrm{s}}

\newcommand{\lqcd}{\Lambda_{\mathrm{QCD}}}
\newcommand{\as}{\alpha_{\mathrm{s}}}

\newcommand{\fig}{Fig.~}
\newcommand{\figs}{Figs.~}
\newcommand{\eq}{Eq.~}
\newcommand{\se}{Sec.~}
\newcommand{\eqs}{Eqs.~}
\newcommand{\re}{Ref.~}
\newcommand{\res}{Refs.~}

\newcommand{\eqdiag}[1]{\raisebox{-5ex}{\includegraphics[height=11ex]{#1}}}

\begin{document}

\author{T. Lappi}
\affiliation{
Department of Physics, %
 P.O. Box 35, 40014 University of Jyv\"askyl\"a, Finland
}

\affiliation{
Helsinki Institute of Physics, P.O. Box 64, 00014 University of Helsinki,
Finland
}
\author{H. M\"antysaari}
\affiliation{
Department of Physics, %
 P.O. Box 35, 40014 University of Jyv\"askyl\"a, Finland
}

\title{
On the running coupling in the JIMWLK equation
}

\pacs{24.85.+p,12.38.Lg}

\begin{abstract}
We propose a new method to implement the running coupling 
constant in the JIMWLK equation, imposing the scale dependence on the correlation 
function of the random noise in the Langevin formulation. 
We  interpret this scale choice as the 
transverse momentum of the emitted gluon in one step of the evolution
and  show that
it is related to the ``Balitsky'' prescription for the
BK equation. This slows down the evolution speed of a 
practical solution of the JIMWLK equation, bringing it closer to 
the $x$-dependence  inferred from fits to HERA data.
We further study our proposal by a numerical comparison of the BK and
JIMWLK equations. 
\end{abstract}

\maketitle

\section{Introduction}

To understand hadronic and nuclear scattering at the high collision energies 
reached at the LHC, one needs a good picture of
high energy behavior of QCD. Much of the phenomenological
work in the field is done within the Color Glass Condensate
(CGC) framework (for reviews see e.g.~\cite{Gelis:2010nm,Lappi:2010ek}).
In this picture, the most convenient choice for the basic degrees of freedom
of a hadronic state are Wilson lines (path ordered exponentials of the 
color field), which describe the eikonal propagation of a high energy 
probe. In the CGC, these are
stochastic variables drawn from an energy-dependent probability distribution.
This energy dependence  is described by the JIMWLK 
renormalization group
equation (see \re\cite{Jalilian-Marian:1997xn,*Jalilian-Marian:1997jx,*Jalilian-Marian:1997gr,%
*Jalilian-Marian:1997dw,*JalilianMarian:1998cb,*Iancu:2000hn,%
*Iancu:2001md,*Ferreiro:2001qy,*Iancu:2001ad,Weigert:2000gi,Mueller:2001uk}),
which can be derived in  weak coupling QCD by successively integrating additional gluonic 
degrees of freedom into the Wilson lines.

With a solution of the JIMWLK equation
one can calculate expectation values for various operators 
formed from the Wilson lines, describing different
scattering processes. For example the simplest one, the dipole,
(see \eq\nr{eq:defdipole}) is related to the inclusive DIS cross section
and to the unintegrated gluon distribution, needed for single inclusive
particle multiplicities.
In phenomenological work, the dipole can most conveniently be obtained from
the Balitsky-Kovchegov (BK)
equation~\cite{Balitsky:1995ub,Kovchegov:1999yj,Kovchegov:1999ua}
which, although the original derivation predates JIMWLK, can now be viewed
as its mean field approximation.
Multiparticle correlation observables, however,
require an understanding of the JIMWLK evolution beyond
the BK approximation. For example, for a calculation
of the azimuthal angle dependence of two-particle 
correlations~\cite{Marquet:2007vb,Albacete:2010pg,Dumitru:2011vk,Dominguez:2011wm,Lappi:2012nh} one needs the expectation values of 
four and six point functions such as the quadrupole,
\eq\nr{eq:defquadrupole}. Another example are long range rapidity correlations, 
e.g. the ``ridge'', which is related to  long range correlations
in the color fields of the colliding 
hadrons~\cite{Dumitru:2008wn,Gelis:2008sz,Dusling:2009ni,Dumitru:2010iy,Dusling:2012ig}.

The inclusion of the running coupling constant in the BK 
equation (rcBK)~\cite{Kovchegov:2006vj,Balitsky:2006wa,Albacete:2007yr} has been an
essential theoretical improvement for phenomenological 
applications~\cite{Albacete:2009fh,Albacete:2010sy,Albacete:2010bs,Kuokkanen:2011je,Albacete:2012xq}.
The running of the QCD coupling slows down the evolution (with $x$ or 
equivalently $\ln \sqrt{s}$) significantly 
from the fixed copupling case, bringing the speed to a rough  agreement
with HERA measurements. Since the full NLO version of the 
equation~\cite{Balitsky:2008zza} has not been extensively applied to phenomenology
(see however recent work in \re\cite{Avsar:2011ds}), rcBK is currently 
the state of the art in phenomenological applications.

The running coupling resums a certain subset of all the NLO corrections to the 
evolution equations. Which subset precisely one chooses to resum is not
uniquely defined, but different choices lead to different ``prescriptions''. Once the
other NLO corrections are included, the physical result should be independent of
this prescription. However,  in present day applications this
is not done yet, and the result will depend on how the running coupling is implemented.
The running coupling constant prescription that is usually used with the BK equation 
is the one derived by Balitsky~\cite{Balitsky:2006wa} (our \eq\nr{eq:bal}), 
since it has been found to minimize
a certain subset of other NLO contributions~\cite{Albacete:2007yr} and provide 
a slow evolution speed roughly consistent with HERA data.

The few existing solutions~\cite{Lappi:2011ju,Dumitru:2011vk}
of the JIMWLK equation with a running coupling have
 not used the same prescription. This is due to the need
to decompose the evolution kernel into a product of two factors
for the numerical solution of the JIMWLK equation.
Instead, what has been used is a ``square root'' running coupling,
\eq\nr{eq:sqrtas}. What we propose in this paper as a solution to overcome this 
technical limitation  is an alternative
form of the running coupling JIMWLK equation, given by 
\eqs\nr{eq:jtimestepsymmeta} and~\nr{eq:newnoisecorr}.

In this paper our starting point is different from 
\res\cite{Kovchegov:2006vj,Balitsky:2006wa,Albacete:2007yr}, where 
one sets out to find the correct modification of the BK kernel, keeping the form
of the BK equation otherwise intact.
Here we want to propose an equation (``rcJIMWLK'')
that keeps intact the functional form of the
JIMWLK equation and there is no particular reason for the resulting
mean field BK approximation to stay exactly the same.
We will, however, argue the BK equation derived from our proposal (\eq\nr{eq:bkfromj1}) 
is for practical
purposes very close to the one with the Balitsky prescription.
Its functional form is also, while different from rcBK, not that different
from structures appearing in the full NLO BK equation.
 We will also compare numerical simulations of both the  BK equation and our proposed
JIMWLK formulation,
starting from the same initial condition.

This paper is organized in the following way. We shall first,
in \se\ref{sec:langevin}, recall the 
Langevin formulation of the JIMWLK equation existing in the literature and
 in \se\ref{sec:fcbk} the derivation of the BK equation from it. We 
proceed  in \se\ref{sec:simple} to  rewrite the Langevin step in a 
simpler way that corresponds to the
the manifestly left-right or ``mirror''~\cite{Kovner:2005jc,Kovner:2005en,Iancu:2011nj}
symmetric form of JIMWLK. We then motivate our
proposal for  a running coupling version of the equation in 
\se\ref{sec:rcj} and discuss its relation to the rcBK equation
in \se\ref{sec:rcbk}. In \se\ref{sec:num} we compare numerical solutions of the
rcBK and rcJIMWLK equations before concluding in \se\ref{sec:conc}.

\section{Langevin formulation of JIMWLK}
\label{sec:langevin}

The  equation for the probability distribution of Wilson lines
can be seen as a functional Fokker-Planck
equation~\cite{Weigert:2000gi}. This in turn can be re-expressed as a
functional Langevin equation for the Wilson lines
themselves~\cite{Blaizot:2002xy},
\begin{equation}\label{eq:langevin1}
\frac{\ud}{\ud y} V_\xt = V_\xt (i t^a) \left[
\int_\zt 
\varepsilon_{\xt,\zt}^{ab,i} \; \xi_\zt(y)^b_i  + \sigma_\xt^a 
\right] .
\end{equation}
Here we denote two dimensional vectors by $\xt,\cdots$ and their lengths as 
$\xtt$.  In the interest of space the coordinate arguments are
written  as subscripts. The Wilson line $V$ is a unitary matrix in the
fundamental
representation, generated by $t^a$  and $i=1,2$ is a transverse spatial index.
The first term in \eq\nr{eq:langevin1} is proportional to a stochastic random noise
and the second one is a deterministic ``drift'' term.
The coefficient of the noise in the stochastic term is
\begin{equation}
 \varepsilon_{\xt,\zt}^{ab,i} = \left(\frac{\as}{\pi}\right)^{1/2}\;
K_{\xt-\zt}^i
\left[1-U_\xt^\dag  U_\zt\right]^{ab},
\label{eq:sqrt}
\end{equation}
which can be thought of the ``square root'' of the  JIMWLK Hamiltonian. 
We  follow here the notation of \re\cite{Dumitru:2011vk} and denote
the adjoint representation Wilson line by $U$; it is related to
the fundamental one by 
\begin{equation}
U_\xt^{ab} = 2 \tr t^a V_\xt t^b V_\xt^\dag.
\end{equation}
We have denoted the coordinate dependent part of the kernel (essentially
the $q \to q g$ splitting wavefunction in LC quantization) by
$K^i_{\xt-\zt}$. 
For the  fixed coupling continuum theory
it is simply
\begin{equation}
K^i_{\xt-\zt} = \frac{ (\xt-\zt)^i }{ (\xt-\zt)^2}; \quad \Kt_\xt = \frac{\xt}{\xtt^2} .
\end{equation}
In the ``square root'' prescription one would modify 
this to account for the running coupling, and on a lattice
one needs to render  $\Kt_{\xt-\zt}$ consistent with periodic boundary 
conditions. For this reason we will for the moment
keep this more general notation.
The noise $\xit = (\xi_1^a t^a,\xi_2^a t^a ) $
is taken to be Gaussian and local in color, transverse coordinate
and rapidity $y$ (evolution time) with
$\langle \xi_\xt(y)^b_i\rangle =0$ and 
\begin{equation}\label{eq:contgauss}
\langle \xi_\xt(y)^a_i \xi_\yt(y')^b_j\rangle = \delta^{ab}
\delta^{ij}\delta^{(2)}_{\xt\yt} \delta(y-y').
\end{equation}
The deterministic term is
\begin{equation}\label{eq:sigma}
\sigma_\xt^a = -i \frac{\as} {2\pi^2}
\int_{\zt} S_{\xt -\zt}
 \widetilde{\tr} \left[ T^a U_\xt^\dag U_\zt\right],
\end{equation}
where $T^a$ is a generator of the adjoint representation,
$\widetilde{\tr}$ the adjoint representation trace
and we have introduced the kernel
\begin{equation}
S_{\xt -\zt} = \frac{1}{ (\xt -\zt)^2 }.
\end{equation}

The Langevin process is both derived~\cite{Blaizot:2002xy} 
and implemented numerically
\cite{Rummukainen:2003ns,Kovchegov:2008mk} in a discrete
evolution time. It is therefore useful to see explicitly what 
happens on one step in rapidity
from $y_n$ to $y_{n+1} = y_n + \ud y$. 
The change in the Wilson line is
\begin{multline}\label{eq:jtimestep1}
V_\xt(y + \ud y) = V_\xt(y) \\ \times
\exp \bigg\{ i t^a \int_\zt 
\varepsilon_{\xt,\zt}^{ab,i} \xi_\zt^{b,i} \sqrt{\ud y}  + 
\sigma_\xt^a \ud y
\bigg\}.
\end{multline}
Note that the rapidity delta function in \eq\nr{eq:contgauss} becomes
a Kronecker delta divided by the magnitude of the timestep
$\delta(y_m-y_n) \to \delta_{m,n}/\ud y$, which we have removed from 
 the normalization of the noise to keep the power counting
in $\ud y$ more explicit. Thus the 
variance of the Gaussian noise is now
\begin{equation}\label{eq:discrgauss}
\langle \xi_\xt(m)^a_i \xi_\yt(n)^b_j\rangle = \delta^{ab}
\delta^{ij}\delta^{(2)}_{\xt\yt} \delta_{mn},
\end{equation}
and the stochastic term in \eq\nr{eq:jtimestep1} is explicitly
proportional to $\sqrt{\ud y}$. The step in rapidity 
$\ud y$ is assumed to be small so that  we can expand
all the quantities on one timestep to the order $\ud y$, i.e.
to second order in the stochastic term and to first order in the
deterministic term.

\section{Deriving BK from JIMWLK at fixed coupling}
\label{sec:fcbk}

The BK equation describes the time evolution of the 
two point function or ``dipole''
\begin{equation}\label{eq:defdipole}
\langle \hat{D}_{\xt,\yt} \rangle = 
 \frac{1}{\nc} \left\langle \tr V^\dag_\xt V_\yt \right\rangle.
\end{equation}
It can be derived from JIMWLK in a mean field approximation, 
where four point functions of Wilson lines are assumed 
to factorize into products of two point functions; what 
matters here is, however, not this truncation of the Balitsky hierarchy
but the relation between the BK and JIMWLK kernels.
Let us start by rederiving the BK equation directly from
\eq\nr{eq:jtimestep1}. In order to get the rapidity 
derivative of $\hat{D}_{\xt,\yt}$ we must 
express $\hat{D}_{\xt,\yt}(y+\ud y)$ as $\hat{D}_{\xt,\yt}(y)$ plus a 
correction proportional to $\ud y$. This is done by inserting 
the timestep \nr{eq:jtimestep1} into the expression for
$\hat{D}_{\xt,\yt}(y+\ud y)$ and expanding to order $\ud y$. 
Since we are, in the BK equation, interested only 
in the evolution of the mean scattering amplitude we
can then take an expectation value over the 
stochastic noise $\xi$, which is independent of the
Wilson lines from the previous steps in rapidity
(i.e. the Langevin equation is discretized with the It\^o interpretation).
One then uses the Fierz identity
\begin{equation}
t^a M t^a = \frac{1}{2}\tr M - \frac{1}{2\nc}M
\end{equation}
to remove the generators remaining from the Gaussian avarage
$\langle \xi^a \xi^b\rangle t^a \cdots t^b \cdots = 
t^{a}\cdots t^a \cdots$.
To clarify the developments in what will follow, let us gather the 
terms in $(\hat{D}_{\xt,\yt}(y+\ud y) - \hat{D}_{\xt,\yt}(y))/\ud y$  here one by one:
\begin{widetext}
\begin{eqnarray}
\nonumber
 \frac{1}{\nc} \tr V^\dag_\xt V_\yt (i \sigma_\yt^a t^a)
&=& \frac{\as}{\pi^2}
\int_{\zt} S_{\yt-\zt} \frac{\nc}{4} \left\{
 \frac{1}{\nc} \tr V^\dag_\xt V_\zt
 \frac{1}{\nc} \tr V^\dag_\zt V_\yt
-
 \frac{1}{\nc} \tr V^\dag_\yt V_\zt
 \frac{1}{\nc} \tr V^\dag_\zt V_\yt V^\dag_\xt V_\yt
\right\}
\\
\nonumber
 \frac{1}{\nc} \tr (-i \sigma_\xt^a t^a) V^\dag_\xt V_\yt 
&=& \frac{\as}{\pi^2}
\int_{\zt} S_{\xt-\zt} \frac{\nc}{4} \left\{
 \frac{1}{\nc} \tr V^\dag_\xt V_\zt
 \frac{1}{\nc} \tr V^\dag_\zt V_\yt
-
 \frac{1}{\nc} \tr V^\dag_\zt V_\xt
 \frac{1}{\nc} \tr V^\dag_\xt V_\yt V^\dag_\xt V_\zt
\right\}
\\
-\frac{1}{2}  \frac{1}{\nc} \tr V^\dag_\xt V_\yt 
 \left\langle \left(\int_\zt  t^a \varepsilon_{\yt,\zt}^{ab,i}  
     \xi_\zt^{b,i} \right)^2 \right\rangle
&=& \frac{\as}{\pi^2}
\int_{\zt} 
\Kt_{\yt-\zt}^2  \frac{\nc}{4} 
\bigg\{
 \frac{1}{\nc} \tr V^\dag_\xt V_\zt
 \frac{1}{\nc} \tr V^\dag_\zt V_\yt
-
 \frac{2}{\nc} \tr V^\dag_\xt V_\yt
\nonumber \\ && \quad \quad\quad
+
 \frac{1}{\nc} \tr V^\dag_\yt V_\zt
 \frac{1}{\nc} \tr V^\dag_\zt V_\yt V^\dag_\xt V_\yt
\bigg\}
\\
\nonumber
-\frac{1}{2} \frac{1}{\nc} 
\tr 
\left\langle \left(\int_\zt  t^a \varepsilon_{\xt,\zt}^{ab,i}  
\xi_\zt^{b,i} \right)^2 \right\rangle 
V^\dag_\xt V_\yt
&=& \frac{\as}{\pi^2}
\int_{\zt} 
\Kt_{\xt-\zt}^2 \frac{\nc}{4} \bigg\{
 \frac{1}{\nc} \tr V^\dag_\xt V_\ut
\ \  \frac{1}{\nc} \tr V^\dag_\zt V_\yt
-
 \frac{2}{\nc} \tr V^\dag_\xt V_\yt
\nonumber \\ \nonumber && \quad \quad \quad 
+
 \frac{1}{\nc} \tr V^\dag_\zt V_\xt
\ \frac{1}{\nc} \tr V^\dag_\xt V_\ut V^\dag_\xt V_\yt
\bigg\}
\\
\nonumber
 \frac{1}{\nc} 
\tr V^\dag_\xt V_\yt
\left\langle  
t^a \varepsilon_{\yt,\zt}^{ab,j} \xi_\zt^{b,j}
 t^c \varepsilon_{\xt,\ut}^{cd,i}  \xi_\ut^{d,i}  \right\rangle
&=& \frac{\as}{\pi^2}
\int_{\zt} 
\Kt_{\xt-\zt} \cdot \Kt_{\yt-\zt} \nc 
\bigg\{
 \frac{1}{\nc} \tr V^\dag_\xt V_\yt
-
 \frac{1}{\nc} \tr V^\dag_\xt V_\zt
 \frac{1}{\nc} \tr V^\dag_\zt V_\yt
\bigg\}.
\end{eqnarray}
The terms that are linear  in the noise vanish in with the expectation value.
Combining the equations above yields
\begin{multline}\label{eq:bkfromlan}
\frac{\hat{D}_{\xt,\yt}(y+\ud y) - \hat{D}_{\xt,\yt}(y)}{\ud y}
=
\frac{\as \nc}{2\pi^2}
\int_{\zt}
\Bigg\{
\left(
\frac{S_{\xt-\zt} + \Kt^2_{\xt-\zt}}{2}
+ \frac{S_{\zt-\yt} + \Kt^2_{\zt-\yt}}{2}
-2 \Kt_{\xt-\zt}\cdot\Kt_{\zt-\yt} 
\right)
 \frac{1}{\nc} \tr V^\dag_\xt V_\zt
 \frac{1}{\nc} \tr V^\dag_\zt V_\yt
\\
+
\frac{1}{2}
\left(
\Kt^2_{\xt-\zt}
- S_{\xt-\zt}
\right)
 \frac{1}{\nc} \tr V^\dag_\zt V_\xt 
 \frac{1}{\nc} \tr V^\dag_\xt V_\zt V^\dag_\xt V_\yt
+ 
\frac{1}{2}
\left(
\Kt^2_{\zt-\yt}
-S_{\zt-\yt}
\right)
 \frac{1}{\nc} \tr V^\dag_\yt V_\zt 
 \frac{1}{\nc} \tr V^\dag_\zt V_\yt V^\dag_\xt  V_\yt
\\
-
\left(
\Kt^2_{\xt-\zt}
+ \Kt^2_{\zt-\yt}
-2 \Kt_{\xt-\zt }\cdot\Kt_{\zt-\yt} 
\right)
 \frac{1}{\nc} \tr V^\dag_\xt V_\yt
\Bigg\}.
\end{multline}
\end{widetext}
A few remarks are in place based on an inspection of \eq\nr{eq:bkfromlan}.
Firstly, one  cannot 
separately modify the kernels $S$ and $\Kt$ in order to 
discretize the theory or to introduce a running coupling. If the 
relation $\Kt^2= S$ is not fulfilled one introduces an
additional six point correlation  structure into the evolution equation of the 
dipole. Although this kind of higher point functions do appear in the NLO BK equation,
they should be corrections at order $\as^2$ and not present here at the leading
order, fixed $\as$ level.
Secondly, the stochastic term (identifiable from the kernel $\Kt$ instead of $S$)
alone gives the correct BK equation virtual term, i.e. 
the one with only the original dipole on the r.h.s.
The stochastic term alone, however, gives an incorrect kernel for the real term and
the additional sextupole term, which are then corrected by the addition
of the deterministic term.
We shall show in the following that the additional complication 
of having to include a deterministic term is an artefact resulting from writing the evolution
step only in terms of a multiplication of the Wilson line from the right (right 
covariant derivative). The JIMWLK equation is in fact 
naturally left-right-symmetric, which can be made explicit 
by expressing it in terms of both left and right 
derivatives~\cite{Iancu:2011nj,Kovner:2005jc,Kovner:2005en}.

The fixed coupling BK equation, included here for reference, can be obtained 
from \eq\nr{eq:bkfromlan} by setting $S=\Kt^2$ to get
\begin{multline}\label{eq:fixedas}
\partial_y \hat{D}_{\xt,\yt}(y) 
=
\frac{\as \nc}{2 \pi^2}
\int_{\zt}
\bigg(
\Kt_{\xt-\zt}^2 
+
  \Kt_{\yt-\zt}^2
-2 \Kt_{\xt-\zt}\cdot\Kt_{\yt-\zt}
\bigg)
\\ \times 
\left[
\hat{D}(\xt,\zt)\hat{D}(\zt,\yt) 
- \hat{D}(\xt,\yt)
\right].
\end{multline}
To obtain the BK equation  for the dipole 
$\langle \hat{D} \rangle$ one then closest this by the mean field
approximation $\langle \hat{D} \hat{D} \rangle \approx \langle \hat{D} \rangle
\langle \hat{D} \rangle$.

The ``square root'' running coupling prescription of 
\res\cite{Lappi:2011ju,Dumitru:2011vk} was simply justified as the 
simplest possible modification of the kernel $\Kt$ to include a dependence
of the coupling on the distance scale. If the only modification possible
is that of the kernel $\Kt$, one cannot impose 
a running of the coupling as a function of the ``parent'' dipole size 
$|\xt-\yt|$. In stead, one is left with the ``square root'' coupling
\begin{multline}\label{eq:sqrtas}
\partial_y \hat{D}_{\xt,\yt}(y) 
=
\frac{\nc}{2 \pi^2}
\int_{\zt}
\bigg(
\as(|\xt-\zt|)  \Kt_{\xt-\zt}^2 
+
\as(|\yt-\zt|)  \Kt_{\yt-\zt}^2
\\
-2 \sqrt{\as(|\xt-\zt|)\as(|\yt-\zt|) }\Kt_{\xt-\zt}\cdot\Kt_{\yt-\zt}
\bigg)
\\ \times 
\left[
\hat{D}(\xt,\zt)\hat{D}(\zt,\yt) 
- \hat{D}(\xt,\yt)
\right].
\end{multline}
The argument of the kernel $\Kt$
is, however, not the only scale in the problem that the coupling can depend on. 
Introducing an $\as$ into the correlation function of the noise is another option
which, as we will argue in \se\ref{sec:rcj}, has a nice physical interpretation.
 But first it is useful to simplify the 
Langevin form of the equation slightly.

\section{Simpler form of the Langevin step}
\label{sec:simple}

The presence of the deterministic term in \eq\nr{eq:langevin1}
is somewhat of an inconvenience for the discussions that will follow. 
It is also a relatively expensive part of the numerical solution
of the equation. As discussed in \re\cite{Mueller:2001uk}, the 
deterministic term is related to the need to transform ``left'' and
``right'' covariant derivatives into each other. In the Langevin
formulation this means that it is a consequence of wanting to write 
the evolution step as a multiplication of the Wilson line only from one
side. We shall now show how one 
can write the rapidity step with only a stochastic term by
allowing for a multiplication of the Wilson line from both left and right.

 In order to derive an equation that
describes the whole Balitsky hierarchy, not just the evolution
for the dipole,  we must start by finding out what 
exactly is the change of a single Wilson line at point $\xt$
to sufficient order in $\ud y$. 
In the evolution step of a generic Wilson line operator
$V^\dag_{\xti} \otimes \cdots \otimes V^\dag_{\xtn} 
\otimes V_{\yti} \otimes \cdots \otimes V_{\ytn}$
one needs to know, in the limit $\ud y\to 0$, its change up to 
$\mathcal{O}(\ud y )$-terms; the higher ones vanish in the continuous
rapidity limit. Now a $\mathcal{O}(\ud y )$-term can come from either the product of 
two $\mathcal{O}\left(\sqrt{\ud y}\right)$-terms from different Wilson lines, or from
a $\mathcal{O}(\ud y )$-term from a single Wilson line. It follows that 
for a single Wilson line we must, in the Langevin timestep,
keep the terms linear in the noise $\xi$, which could be contracted
with the noise term coming from another Wilson line. One must also keep the
$\mathcal{O}(\ud y )$ terms, but for them one can take the expectation value 
of the noise term, because any correlation between them and the evolution
of another Wilson line would be higher order in $\ud y$.
Expanding \eq\nr{eq:jtimestep1} to order $\ud y$ and taking
the expectation value of the quadratic noise term one gets
\begin{multline}\label{eq:jtimestepexp}
V_\xt(y + \ud y) = V_\xt(y) \bigg[
1 
\\
+ i \frac{\sqrt{\as \ud y }}{\pi}\int_\zt \Kt_{\xt-\zt} \cdot \left( \xit_\zt 
- V^\dag_\xt V_\zt \xit_\zt V^\dag_\zt V_\xt \right)
\\
-
\frac{\as \ud y }{2 \pi^2}
\int_\zt \Kt^2_{\xt-\zt}
\left(2 - U^\dag_\xt U_\zt  - U^\dag_\zt U_\xt \right)^{ab} t^a t^b
\\
+ 
\frac{\as \ud y }{2 \pi^2}
\int_\zt S_{\xt-\zt}
\left(U^\dag_\xt U_\zt  - U^\dag_\zt U_\xt \right)^{ab} t^a t^b
\bigg].
\end{multline}
Here $ V^\dag_\xt V_\zt \xit_\zt V^\dag_\zt V_\xt  $ is simply the
adjoint representation product $t^a (U^\dag_\xt U_\zt)^{ab} \xit^b_\zt$
written out in the fundamental representation. The terms coming from 
the noise and deterministic contributions can be identified from 
the kernels $\Kt^2$ and $S$ respectively. The effect of the 
deterministic term is to cancel $U^\dag_\zt U_\xt $  and
double the other term $U^\dag_\xt U_\zt$. Now setting
$S$ equal to $\Kt^2$ and also expressing
the remaining adjoint Wilson lines in terms of fundamental 
ones we can write  this as
\begin{multline}\label{eq:jtimestepexp2}
V_\xt(y + \ud y) = V_\xt
\\
+ i \frac{\sqrt{\as \ud y }}{\pi}\int_\zt \Kt_{\xt-\zt} 
\cdot( V_\xt \xit_\zt  
-  V_\zt \xit_\zt V^\dag_\zt V_\xt  )
\\
-
\frac{\as \ud y }{2 \pi^2}
\int_\zt \Kt^2_{\xt-\zt}
(V_\zt \tr V^\dag_\zt V_\xt- \nc V_\xt).
\end{multline}
Up to this order in $\ud y$ this is equivalent to the re-exponentiated form 
\begin{multline}\label{eq:jtimestepsymm}
V_\xt(y + \ud y) = 
\exp\left\{-i\frac{\sqrt{\as \ud y }}{\pi}\int_\zt 
  \Kt_{\xt-\zt} \cdot ( V_\zt \xit_\zt V^\dag_\zt) \right\}
\\
\times
V_\xt 
\exp\left\{i\frac{\sqrt{\as \ud y }}{\pi}\int_\zt 
  \Kt_{\xt-\zt} \cdot \xit_\zt \right\},
\end{multline}
where we have again, in the order $\ud y$ replaced 
$\langle \xi^2\rangle$ by $\xi^2$, the difference
being higher order in $\ud y$.

Equation~\nr{eq:jtimestepsymm}, to be compared to the earlier form 
\nr{eq:jtimestepexp}, is the main result of this section.
Now the Wilson line is multiplied 
from both the left and the right, and consequently there is no more
drift term. 
In addition to having fewer separate terms to calculate,
\eq\nr{eq:jtimestepsymm}
is significantly  faster to evaluate numerically, because it 
avoids the need to reconstruct the adjoint representation Wilson line in
\eq\nr{eq:sigma}. We stress that \eq\nr{eq:jtimestepsymm}  is
 precisely equivalent to the Langevin step derived
in \cite{Blaizot:2002xy} in the limit of continuous evolution time
$\ud y \to 0$. The difference comes only in terms at order 
$\mathcal{O}\left(\sqrt{\ud y}^3\right)$,
which are neglected both here and in the derivation of the original equation 
in Ref.~\cite{Blaizot:2002xy}.

Note that the multiplication by the Wilson line at the coordinate $\zt$
from the left is a choice related to the form of the original choice 
 to use a right derivative in \nr{eq:langevin1}. One could easily move 
it to the right by defining a rotated noise
\begin{equation}
\xitilt_\zt = \xitilt^a_\zt t^a
=  V_\zt \xit_\zt V^\dag_\zt.
\end{equation}
It is easy to see that the distribution of $\xitilt_\zt$ is 
the same as that of $\xit_\zt$, namely
\begin{equation}\label{eq:discrgausstil}
\langle \widetilde{\xi}_\xt^{a,i} \widetilde{\xi}_\yt^{b,j}\rangle 
= \delta^{ab} \delta^{ij}\delta^{(2)}_{\xt-\yt}.
\end{equation}
In terms of  $\xitilt_\zt$ the timestep \nr{eq:jtimestepsymm} is 
\begin{multline}\label{eq:jtimestepsymm2}
V_\xt(y + \ud y) = \exp\left\{-i\frac{\sqrt{\as \ud y }}{\pi}\int_\zt 
  \Kt_{\xt-\zt} \cdot \xitilt_\zt
\right\}
\\
\times
V_\xt 
\exp\left\{i\frac{\sqrt{\as \ud y }}{\pi}\int_\zt 
  \Kt_{\xt-\zt} \cdot
( V^\dag_\zt  \xitilt_\zt V_\zt) 
 \right\}.
\end{multline}

\section{Setting the scale of the coupling}
\label{sec:rcj}

\begin{figure*}
\centerline{
\includegraphics[width=0.3\textwidth]{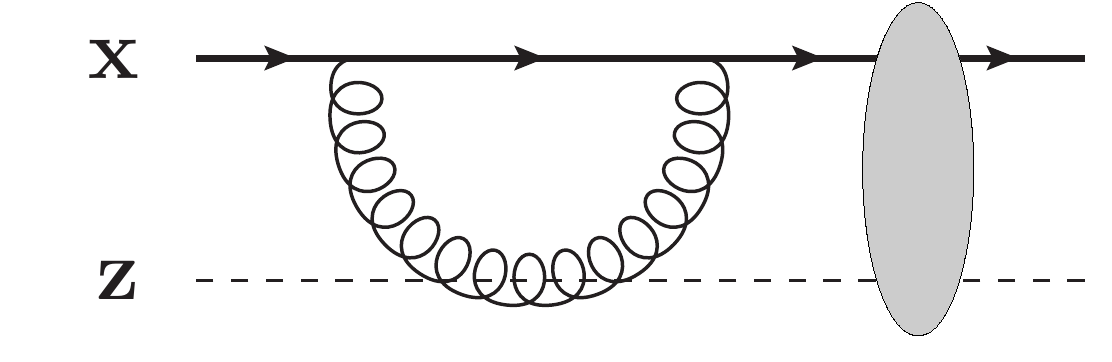}
\includegraphics[width=0.3\textwidth]{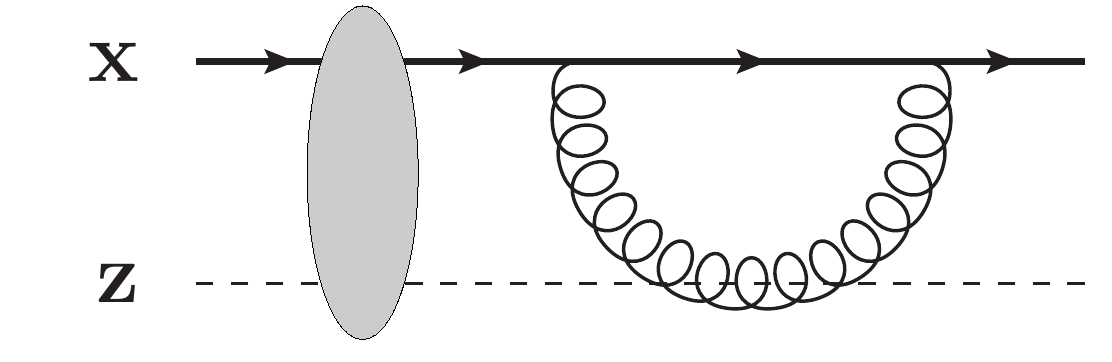}
\includegraphics[width=0.3\textwidth]{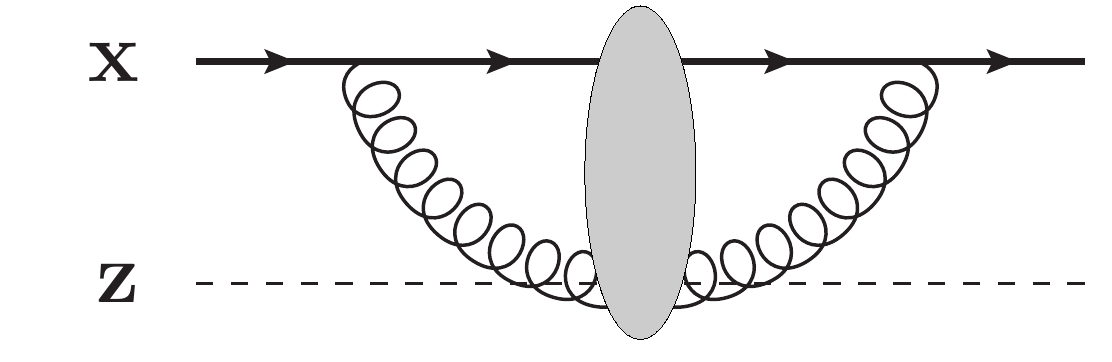}}
\caption{\label{fig:virt}
Virtual diagrams.
}
\end{figure*}
\begin{figure*}
\centerline{
\includegraphics[width=0.3\textwidth]{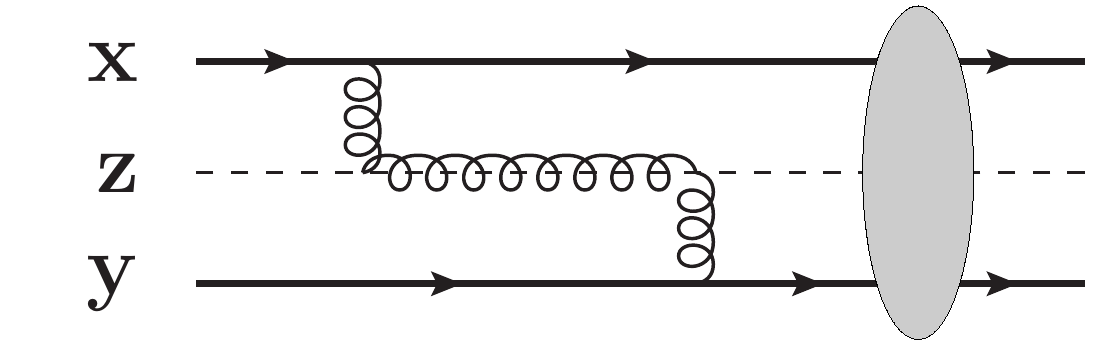}
\includegraphics[width=0.3\textwidth]{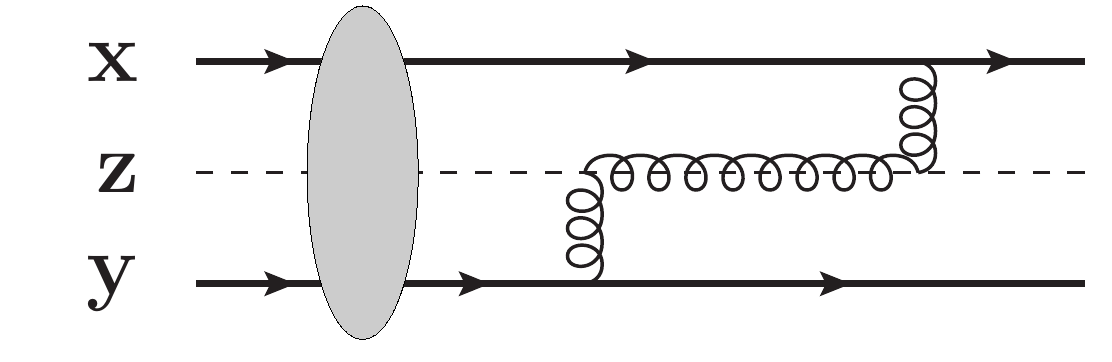}
\includegraphics[width=0.3\textwidth]{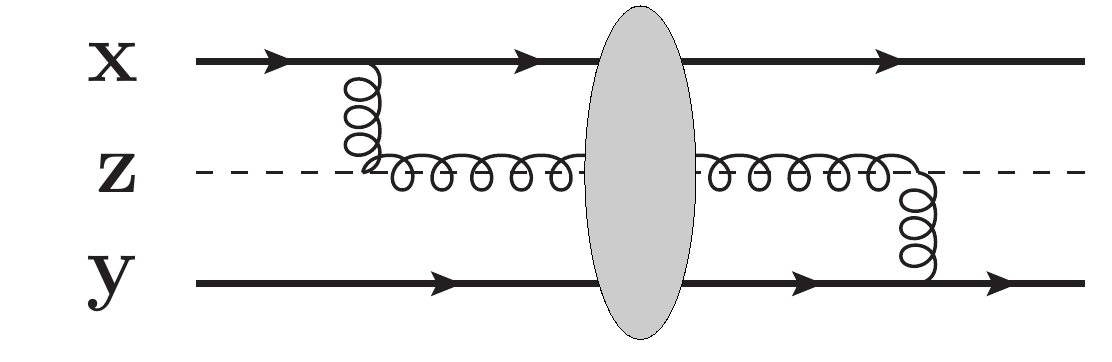}}
\caption{\label{fig:real}
Real diagrams.
}
\end{figure*}

\begin{figure*}
\begin{eqnarray*}
\int_{\zt}
\eqdiag{virta}
&\to& 
\int_{\zt \ztp} 
\int \frac{\ud^2\kt}{(2\pi)^2}
e^{i\kt \cdot ( \zt -\ztp)}
\eqdiag{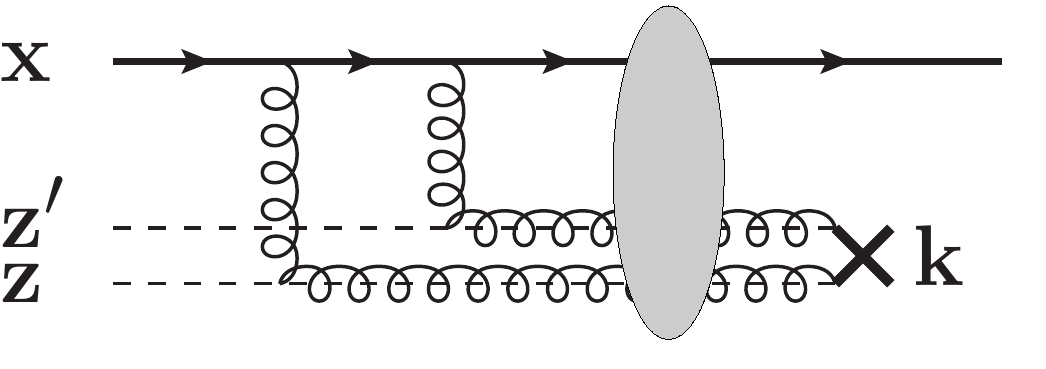}
\\
\int_{\zt}
\eqdiag{virtb}
&\to& 
\int_{\zt \ztp} 
\int \frac{\ud^2\kt}{(2\pi)^2}
e^{i\kt \cdot ( \zt -\ztp)}
\eqdiag{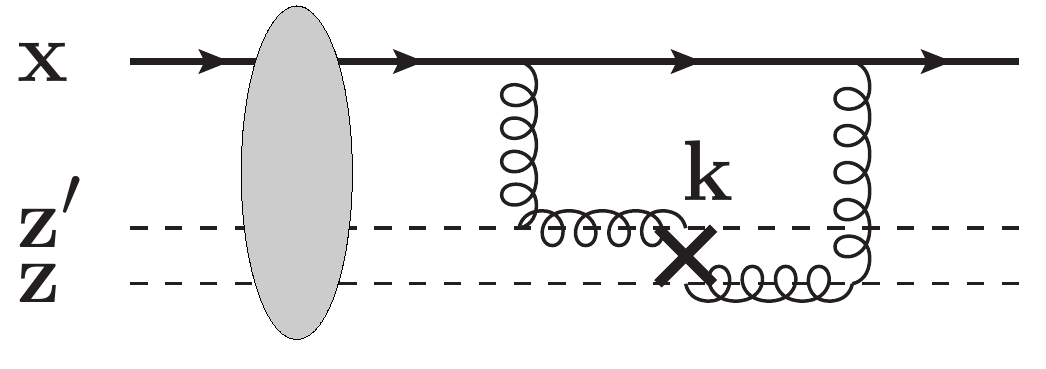}
\\
\int_{\zt}
\eqdiag{virtc}
&\to& 
\int_{\zt \ztp} 
\int \frac{\ud^2\kt}{(2\pi)^2}
e^{i\kt \cdot ( \zt -\ztp)}
\eqdiag{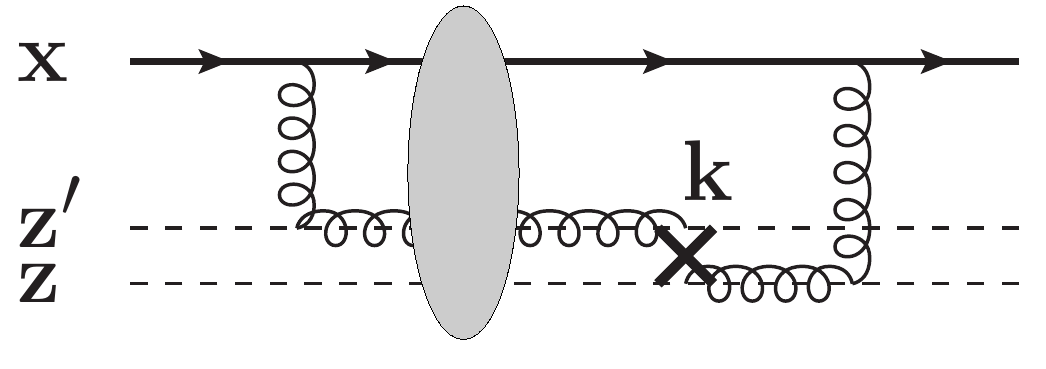}
\end{eqnarray*}
\caption{Virtual diagrams in \fig\ref{fig:virt} rewritten with the 
explicit momentum argument used as the scale of the running coupling.}
\label{fig:virtampli}
\end{figure*}

\begin{figure*}
\begin{eqnarray*}
\int_{\zt}
\eqdiag{reala}
&\to& 
\int_{\zt \ztp} 
\int \frac{\ud^2\kt}{(2\pi)^2}
e^{i\kt \cdot ( \zt -\ztp)}
\eqdiag{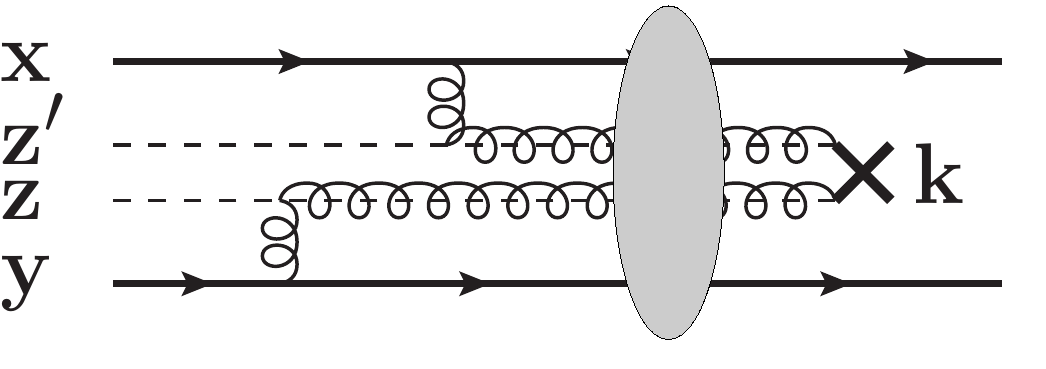}
\\
\int_{\zt}
\eqdiag{realb}
&\to& 
\int_{\zt \ztp} 
\int \frac{\ud^2\kt}{(2\pi)^2}
e^{i\kt \cdot ( \zt -\ztp)}
\eqdiag{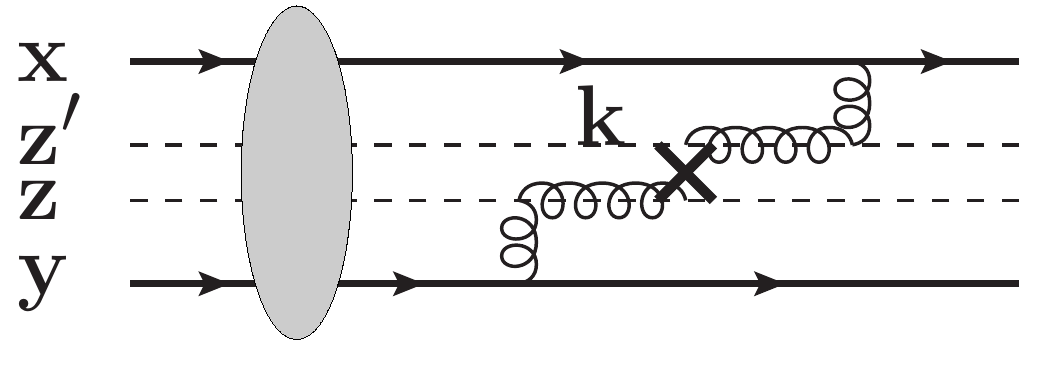}
\\
\int_{\zt}
\eqdiag{realc}
&\to& 
\int_{\zt \ztp} 
\int \frac{\ud^2\kt}{(2\pi)^2}
e^{i\kt \cdot ( \zt -\ztp)}
\eqdiag{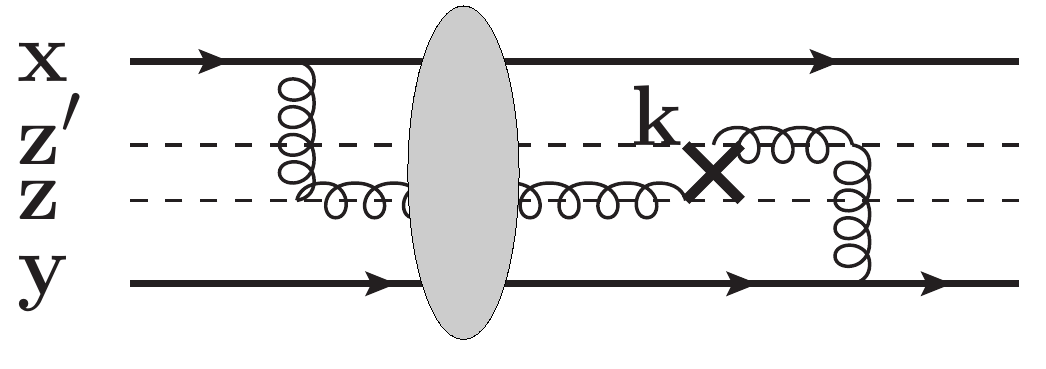}
\end{eqnarray*}
\caption{Real diagrams in \fig\ref{fig:real}  rewritten with the 
explicit momentum argument used as the scale of the running coupling.}
\label{fig:realampli}
\end{figure*}

Recall that the change in a general correlator of $n$ Wilson lines in
one infinitesimal step in rapidity is obtained by changing all the Wilson lines
by \eq\nr{eq:jtimestepsymm}, developing to order $\xit^2$ (i.e. to order
$\ud y$) and taking the expectation values with the probability distribution 
of the noise $\xit$. Physically keeping only the quadratic order in $\xit$ 
corresponds to the fact that JIMWLK is a LO evolution equation, derived
by considering the emission of only one gluon. The interpretation
of the contractions between two $\xit$'s is that one corresponds
to the emission of the gluon in the amplitude and the other one 
in the complex conjugate.
In the formulation that includes the deterministic term this 
interpretation is not as straightforward, since in \eq\nr{eq:langevin1}
one is summing a noise term corresponding to a gluon emission amplitude
and the deterministic term which is an emission probability, i.e. amplitude squared.

In order to have a better picture of the physics contained in one evolution step
we shall now describe, adapting the ``simple derivation of the JIMWLK
equation'' of~\re\cite{Mueller:2001uk}, the contractions of 
the $\xit$'s in \eq\nr{eq:jtimestepsymm} in a diagrammatic notation.
We want to apply the evolution step \eq\nr{eq:jtimestepsymm} to calculate
the change of  a generic Wilson line operator
$V^\dag_{\xti} \otimes \cdots \otimes V^\dag_{\xtn} 
\otimes V_{\yti} \otimes \cdots \otimes V_{\ytn}$
in an infinitesimal step in rapidity. 
We now have to distinguish between several
different cases of how these contractions are done, drawn in \figs\ref{fig:virt}
and~\ref{fig:real}:
\begin{itemize}
\item The square of the noise multiplying the Wilson line from the left corresponds
to the emission and reasorption of a gluon before interaction with the target. 
The noise term $\xit_\zt$ is color rotated into $V_\zt \xit_\zt V^\dag_\zt$ 
in \eq\nr{eq:jtimestepsymm}, but in the 
square  the color rotation cancels. Thus 
we can express this contribution as the left diagram in \fig\ref{fig:virt}.
\item Similarly we can take the square of the $\xit$ multiplying the Wilson 
line from the right. Now the noise is not color rotated, i.e. this corresponds
to emission from and absorption by the Wilson line after the target, shown in the middle
diagram of \fig\ref{fig:virt}.
\item We can also contract the $\xit$ multiplying the Wilson line from the left 
with the one multiplying it from the right. Physically this is equivalent to
the interference between gluon emission before and after the interaction with the 
target. In the diagrammatic notation this is expressed as a gluon being emitted  
before the target, propagating though it, and being reabsorbed, drawn on the right in 
\fig\ref{fig:virt}.
\item One can also have a contraction between the $\xit$'s from the step
of one Wilson line at coordinate $\xt$ with another one at $\yt$. If both 
$\xit$'s multiply the Wilson lines from the left (or the hermitean conjugates from the 
right), they correspond to the diagram where the gluon is emitted and absorbed before
the target, the color rotations again cancel.  This contribution
is shown on the left of  \fig\ref{fig:real}.
\item If the contracted $\xit$'s are the ones to the right of the Wilson lines, 
we have an interference between gluon emission from two different Wilson lines after the target,
as shown in the middle of \fig\ref{fig:real}.
\item Finally we can have an interference between the emission of a gluon from the Wilson line at
$\xt$ before the collision and from another one at $\yt$ after the collision, as shown 
on the right of \fig\ref{fig:real}. This corresponds to the contraction between one $\xit$ on the
left and another one on the right of a Wilson line.
\end{itemize}

If now $\xit_\ut$ corresponds to the emission of a gluon at
coordinate $\ut$ and the other $\xit_\vt$ that it is to be contracted
with is interpreted as the absorption of the same gluon, what
is the physical meaning of the delta function 
$\delta^2(\ut-\vt)$ in the correlation function of the noise?
The answer to this question that we propose in this paper
lies in realizing that in fact the 
``absorption'' of the gluon actually corresponds to the complex
conjugate of the amplitude for gluon emission. Thus if the transverse 
momentum of the emitted gluon is $\kt$, the amplitude
is $\sim e^{i\kt\cdot \ut}$ and the complex conjugate 
$\sim e^{-i\kt\cdot \vt}$. The delta function $\delta^2(\ut-\vt)$
is then a result of integrating over all possible transverse momenta
of the emitted gluon in one step of the evolution:
\begin{equation}
\delta^2(\ut-\vt) = \int \frac{\ud^2 \kt}{(2\pi)^2}
e^{i \kt\cdot(\ut-\vt)}.
\end{equation}

The interpretation of the diagrams of  \figs\ref{fig:virt} and~\ref{fig:real}
in terms of the transverse momentum of the gluon
is shown in  \figs\ref{fig:virtampli} and~\ref{fig:realampli}.
Note that all the momenta $\kt$ are taken after the interaction with the 
target, not  before. This corresponds to the choice made to use 
the noise $\xit$ and not the rotated one $\xitilt$.
The integration over the final state gluon momentum
has to be done because we wish to include \emph{all} the degrees of 
freedom in the rapidity interval from $y$ to $y+\ud y$ into an
effective theory description of the target at the new 
rapidity scale $\ud y$. 

What then happens if we wish to include the running coupling constant?                    
We will here rely on the general argument in gauge theory that the
beta function can be computed by considering only corrections to 
the gluon propagator. This is in fact the property that is used
in~\cite{Kovchegov:2006vj,Balitsky:2006wa} to derive the running coupling
constant in BK which is done by summing one loop corrections to the 
gluon propagator. If the running coupling corrections can be derived
from just the gluon propagator, it is natural that the scale 
of the running coupling should be the momentum of the emitted gluon.
This leads directly to our proposal for implementing the running
coupling in the JIWMLK equation, which is to regard the coupling constant as
a property of the correlator of the noise. We then propose to simply
replace the fixed coupling correlator
\begin{equation}
\as \left\langle \xi_\xt^{a,i} \xi_\yt^{b,j} \right\rangle = \delta^{ab}
\as
\delta^{ij} \int \frac{\ud^2 \kt}{(2\pi)^2}
e^{i \kt\cdot(\xt-\yt)}
\end{equation}
by 
\begin{equation}\label{eq:newcorr}
\left\langle \eta_\xt^{a,i} \eta_\yt^{b,j} \right\rangle = \delta^{ab}
\delta^{ij} \int \frac{\ud^2 \kt}{(2\pi)^2}
e^{i \kt\cdot(\xt-\yt)} \as(\kt),
\end{equation}
in terms of which the rapidity step of the Wilson line is now
\begin{multline}\label{eq:jtimestepsymmeta}
V_{\xt}(y + \ud y) = 
\exp\left\{-i\frac{\sqrt{\ud y }}{\pi}\int_\ut 
  \Kt_{\xt-\ut} \cdot ( V_{\ut} \etat_{\ut} V^\dag_{\ut}) \right\}
\\
\times
V_{\xt}
\exp\left\{i\frac{\sqrt{\ud y }}{\pi}\int_\vt 
  \Kt_{\xt-\vt} \cdot \etat_{\vt} \right\},
\end{multline}
with $\etat = ( \eta_1^at^a, \eta_2^a t^a)$.
Note that since the noise correlator is not local in transverse coordinate,
we have broken the left-right symmetry of the fixed 
coupling equation. The physical interpretation of this
is that the symmetry corresponds to time reversal, and using
the momentum of a gluon in the final state after the target 
(as opposed to an initial state one) 
breaks the time reversal symmetry. While this loss of symmetry
in inaesthetic, it seems inevitable at some point at least if
one wants to generalize the evolution to more exclusive 
observables (as discussed e.g. in \re\cite{Marquet:2010cf}).

Due to the FFT algorithm~\cite{Cooley:1965zz} the 
Fourier-transform is a  relatively straightforward and inexpensive 
procedure on the lattice. Thus implementing the correlator
\nr{eq:newcorr} induces only a minor modification to the numerical
algorithm~\cite{Rummukainen:2003ns} used to solve the JIMWLK equation.
In the following we shall denote the new noise correlator by
\begin{equation}\label{eq:newnoisecorr}
\astil_{\xt-\yt} \equiv \int \frac{\ud^2 \kt}{(2\pi)^2}
e^{i \kt\cdot(\xt-\yt)} \as(\kt).
\end{equation}
We emphasize that the notation $\astil_{\xt-\yt}$ does \emph{not}
mean the coupling evaluated at the momentum scale $1/|\xt-\yt|$, but 
is indeed the Fourier transform of the coupling at the scale $\ktt$. Since the 
coupling is a very smooth, logarithmically varying function of $\ktt$, 
the transform $\astil_{\xt-\yt}$ is very sharply peaked around $\xt=\yt$.

\section{Recovering the BK equation}
\label{sec:rcbk}
Let us now calculate the change in the dipole operator during one rapidity step of the 
evolution. This is done by inserting \eq\nr{eq:jtimestepsymmeta} into the expression for
$\hat{D}_{\xt,\yt}(y+\ud y)$, developing in powers of the noise term up to 
$\mathcal{O}(\etat^2)$ and taking the expectation values with the correlator
\nr{eq:newnoisecorr}. The result is
\begin{widetext}
\begin{multline}\label{eq:bkfromj1}
\frac{\hat{D}_{\xt-\yt}(y+\ud y) - \hat{D}_{\xt-\yt}(y)}{\ud y}
=
\frac{\nc}{2\pi^2}
\int_{\ut,\vt} \astil_{\ut-\vt}
\Bigg\{
\left(
\Kt_{\xt-\ut}\cdot\Kt_{\xt-\vt} + \Kt_{\yt-\ut}\cdot\Kt_{\yt-\vt}
-2 \Kt_{\xt-\ut}\cdot\Kt_{\yt-\vt}
\right)
\\ \times 
\frac{1}{2}\left[
\hat{D}_{\xt,\ut}\hat{D}_{\ut,\yt} + 
	\hat{D}_{\xt,\vt}\hat{D}_{\vt,\yt}
- \hat{D}_{\xt,\yt}
- \hat{D}_{\vt,\ut}\hat{Q}_{\xt,\vt,\ut,\yt}
\right],
\end{multline}
\end{widetext}
where the quadrupole operator is
\begin{equation}\label{eq:defquadrupole}
\hat{Q}_{\xt,\vt,\ut,\yt} = \frac{1}{\nc} \tr V^\dag_\xt V_\vt V^\dag_\ut V_\yt.
\end{equation}
From this form it is easy to check that in the fixed coupling case
$\astil{}_{\ut-\vt}= \as \delta^2(\ut-\vt)$ we recover the usual BK equation.

Now the evolution equation for the two point function does not have the same  
functional form as in the fixed coupling case. Our definition of the 
scale of the running coupling as the momentum of the emitted final state gluon 
makes the position of the gluon in transverse coordinate space different in the amplitude
and the complex conjugate, resulting in a more complicated
the structure for the equation.
From the point of view of the full set of NLO corrections to the BK 
equation~\cite{Balitsky:2008zza} 
this is not  unnatural. The NLO BK equation also necessarily involves
 higher point dipole
operators, and the definition of what subset of these corrections 
one chooses to call ``running coupling'' corrections is not unique. 
What is done in the derivations of the running coupling BK 
equation~\cite{Kovchegov:2006vj,Balitsky:2006wa} 
is that one sets out to calculate a certain subset 
of these NLO corrections, namely those that only modify the BK~kernel,
but leave the Wilson line operator part of the BK equation intact.
This particular set of NLO corrections
is not necessarily one that would leave the functional form of the JIMWLK equation 
intact. What we are proposing here, on the other hand, is running coupling JIMWLK equation, 
namely an evolution equation which includes running coupling corrections, but leaves
the functional form of the JIMWLK equation intact. There is no particular reason 
why  this equation should be equivalent to a particular form of the running coupling BK 
equation, and indeed it is not. However, as we will show in the following, it does 
in fact lead to the scale of the coupling being determined by parametrically the same 
length scale in the parts of phase space that dominate the evolution.

In order to see the relation to rcBK equations in the literature in more detail 
we will now try to connect our \eq\nr{eq:bkfromj1}, where the coupling is a function
of a \emph{momentum} scale to one where it depends on the inverse of a transverse 
\emph{coordinate} distance. We shall here use the same notation for both,
$\as(\kt)$ and $\as(\rt)$, while deferring  the explicit expressions
to \se\ref{sec:num}. We will also not attempt to perform the Fourier-transform
from momentum to coordinate space precisely, but merely attempt to identify the dominant
momentum scales in parametrically different parts of the phase space.

Let us denote $\zt = (\ut+\vt)/2$ and $\Delta \zt = (\ut-\vt)$. 
Using the fact that $\astil_{\ut-\vt}$ is very peaked around 
$\ut=\vt$ we then approximate
$\ut\approx \vt \approx \zt$ (i.e. set $\Delta\zt=0$) 
in the part of \nr{eq:bkfromj1} containing the Wilson line operators
\begin{multline}\label{eq:bkwlines}
\frac{1}{2}\left[
\hat{D}_{\xt,\ut}\hat{D}_{\ut,\yt} + 
	\hat{D}_{\xt,\vt}\hat{D}_{\vt,\yt}
- \hat{D}_{\xt,\yt}
- \hat{D}_{\vt,\ut}\hat{Q}_{\xt,\vt,\ut,\yt}
\right]
\\
\overset{\ut=\vt=\zt}{\longrightarrow}
\hat{D}_{\xt,\zt}\hat{D}_{\zt,\yt} 
- \hat{D}_{\xt,\yt},
\end{multline}
which is the same form as in the BK equation.

Turning then to the kernel, we can use
the momentum representation of $\Kt$
\begin{equation}
\Kt_\xt = i(2\pi)^2 \int \frac{\ud^2\kt}{(2\pi)^2} 
\frac{e^{-i \kt\cdot \xt}}{\kt^2}
\end{equation}
to write
\begin{multline}\label{eq:kernelmassage}
\int_{\Delta \zt} \astil_{\Delta \zt}
\Kt_{\xt-\ut}\cdot\Kt_{\yt-\vt}
\\= \frac{1}{(2\pi)^2} \int_{\pt,\qt} 
\as\left(\frac{\pt+\qt}{2}\right)
\frac{\pt\cdot \qt}{\pt^2\qt^2} 
e^{-i\pt\cdot \xt + i \qt\cdot \yt + i (\pt-\qt)\cdot\zt}
\\= 
\frac{1}{(2\pi)^2} \int_{\Pt,\kt} \as(\Pt)
e^{-i\Pt\cdot(\xt-\yt) + i \kt\cdot(\zt-\frac{\xt+\yt}{2})}
\\ \times
\frac{(\Pt+\kt/2)\cdot (\Pt-\kt/2)}{(\Pt+\kt/2)^2(\Pt-\kt/2)^2} 
\end{multline}
where we have changed variables from $\pt,\qt$ to $\Pt = (\pt+\qt)/2$
and $\kt=\pt-\qt$.
Using this form we can write the kernel of \eq\nr{eq:bkfromj1} as
\begin{multline}\label{eq:kernelmassage2}
\int_{\Delta \zt} \astil_{\Delta \zt}
\left(
\Kt_{\xt-\ut}\cdot\Kt_{\xt-\vt} + \Kt_{\yt-\ut}\cdot\Kt_{\yt-\vt}
-2 \Kt_{\xt-\ut}\cdot\Kt_{\yt-\vt}
\right)
\\
= 
\frac{1}{(2\pi)^2} \int_{\Pt,\kt} \as(\Pt)
\frac{(\Pt+\kt/2)\cdot (\Pt-\kt/2)}{(\Pt+\kt/2)^2(\Pt-\kt/2)^2} 
\\
\left[
e^{ i \kt\cdot(\zt-\xt)}
+ e^{ i \kt\cdot(\zt-\yt)}
-2 e^{-i\Pt\cdot(\xt-\yt) + i \kt\cdot(\zt-\frac{\xt+\yt}{2})}
\right].
\end{multline}
We now look separately at two different kinematical regimes. 

\paragraph{Small daughter dipole} Let us first assume 
that one of the daughter dipoles is smaller, e.g. 
$|\xt-\zt| \ll |\yt-\zt| \approx |\xt-\yt|$. Now the dominant term 
in the $\Pt$ and $\kt$-integral is the one that oscillates most slowly, namely
as $e^{ i \kt\cdot(\zt-\xt)}$. Keeping only this term 
we can return to the variables
$\pt$ and $\qt$ and write it as
\begin{equation}
\frac{1}{(2\pi)^2} \int_{\pt,\qt} \as\left(\frac{\pt+\qt}{2}\right)
\frac{\pt\cdot \qt}{\pt^2\qt^2} 
e^{-i(\pt-\qt)\cdot (\xt -\zt)}.
\end{equation}
The exponential factor now makes both $\ptt$ and $\qtt$ of the order
$1/|\xt-\zt|$. We thus approximate $\as((\pt+\qt)/2)\approx \as(|\xt-\zt|)$,
leading us to the estimate
\begin{equation}\label{eq:smalld1}
\sim \frac{\as(\xt-\zt)}{(\xt-\zt)^2}, \quad |\xt-\zt| \ll  |\yt-\zt| \approx |\xt-\yt|
\end{equation}
for the kernel in this limit, and similarly
\begin{equation}\label{eq:smalld2}
\sim \frac{\as(\yt-\zt)}{(\yt-\zt)^2}, \quad |\yt-\zt| \ll  |\xt-\zt| \approx |\xt-\yt|
\end{equation}
for the other small daughter dipole limit.
This is almost an inevitable result, which can also be seen noting that already
at fixed coupling the kernel in this kinematical limit is dominated by the
single term $\Kt_{\xt-\zt}^2$ or $\Kt_{\yt-\zt}^2$,
 and the coupling must depend on the only distance
scale available for it, namely the daughter dipole size.

\paragraph{Small parent dipole} Assume then that
the parent dipole is smaller, e.g. 
$|\xt-\yt| \ll |\xt-\zt| \approx |\yt-\zt|$. We can then approximate
\begin{multline}\label{eq:smallpar}
e^{ i \kt\cdot(\zt-\xt)}
+ e^{ i \kt\cdot(\zt-\yt)}
-2 e^{-i\Pt\cdot(\xt-\yt) + i \kt\cdot(\zt-\frac{\xt+\yt}{2})}
\\
\approx
2e^{ i \kt\cdot(\zt-\frac{\xt+\yt}{2})}
\left(1
- e^{-i\Pt\cdot(\xt-\yt) }
\right),
\end{multline}
from which we immediately see that the scale of the coupling $\Pt$ is 
given by the parent dipole size $\xt-\yt$. The physical intepretation of this 
is that the emission of the additional gluon happens coherently from the
dipole (as evidenced by the fact that the kernel vanishes at $\xt=\yt$, 
reflecting color neutrality), thus
the parent dipole determines the scale of the coupling. 
After changing the scale of the running coupling to $\xt-\yt$ we can then
pull it outside of the $\Pt,\kt$-integral~\nr{eq:kernelmassage2} and
restore the exponential functions to the original form on the
 l.h.s. of \eq\nr{eq:smallpar}.
Doing this we  get the estimate
\begin{equation}\label{eq:smallp}
\sim \as(\xt-\yt)\frac{(\xt-\yt)^2}{(\xt-\zt)^2(\yt-\zt)^2},
 \quad |\xt-\yt| \ll  |\xt-\zt| \approx |\yt-\zt|
\end{equation}
for the kernel.

It is instructive to compare the estimate
\nr{eq:smalld1}, \nr{eq:smalld2} and \nr{eq:smallp} to the ``Balitsky'' 
rcBK equation, which reads, in our notation:
\begin{multline}\label{eq:bal}
\partial_y \hat{D}_{\xt,\yt}(y) 
=
\frac{\nc}{2\pi^2}
\int_{\zt}
\bigg(
\frac{\as(\xt-\zt)\as(\xt-\yt)}{\as(\yt-\zt)}  \Kt_{\xt-\zt}^2 
\\
+
\frac{\as(\yt-\zt)\as(\xt-\yt)}{\as(\zt-\zt)}  \Kt_{\yt-\zt}^2
 \\
-2 \as(\xt-\yt)\Kt_{\xt-\zt}\cdot\Kt_{\yt-\zt}
\bigg)
\\ \times 
\left[
\hat{D}(\xt,\zt)\hat{D}(\zt,\yt) 
- \hat{D}(\xt,\yt)
\right].
\end{multline}
One immediately recognizes that the scale of the coupling in the different
kinematical limits is the same.

As a summary, showing the approximate equivalence of our rcJIMWLK equation, defined by 
\eqs\nr{eq:jtimestepsymmeta} and \nr{eq:newnoisecorr}, 
to the ``Balitsky'' rcBK equation~\nr{eq:bal} commonly used in phenomenology
involved the following steps.
We first derived an equation for the two point function \eq\nr{eq:bkfromj1}
that involved 4 transverse coordinates and a six-point function 
$\hat{D}\hat{Q}$. We subsequently took the limit of only 3 coordinates, in stead of 4, 
for the Wilson line operators, reducing them to the BK form. We then
identified the dominant momentum scales for the coupling in the kernel of
\eq\nr{eq:bkfromj1} in the limits of small and large daughter dipoles compared
to the parent dipole, and 
 interchanged the momentum and the corresponding coordinate distance as the argument
of the running coupling. 
In addition, our \eqs\nr{eq:bkfromj1} and~\nr{eq:bal}
have been written as operator equations corresponding to the first equation
of the infinite Balitsky hierarchy. While the JIMWLK equation solves the whole hierarchy
simultaneously, a solution of just the BK equation requires one to truncate
the hierarchy  by replacing the operators $\hat{D}$ by their expectation values. 
To quantify the effect of these approximations we shall now proceed to a comparison
between numerical evaluations of the BK and JIMWLK equations.

\section{Numerical comparison}
\label{sec:num}

\begin{figure}[t]
\centerline{\includegraphics[width=0.45\textwidth]{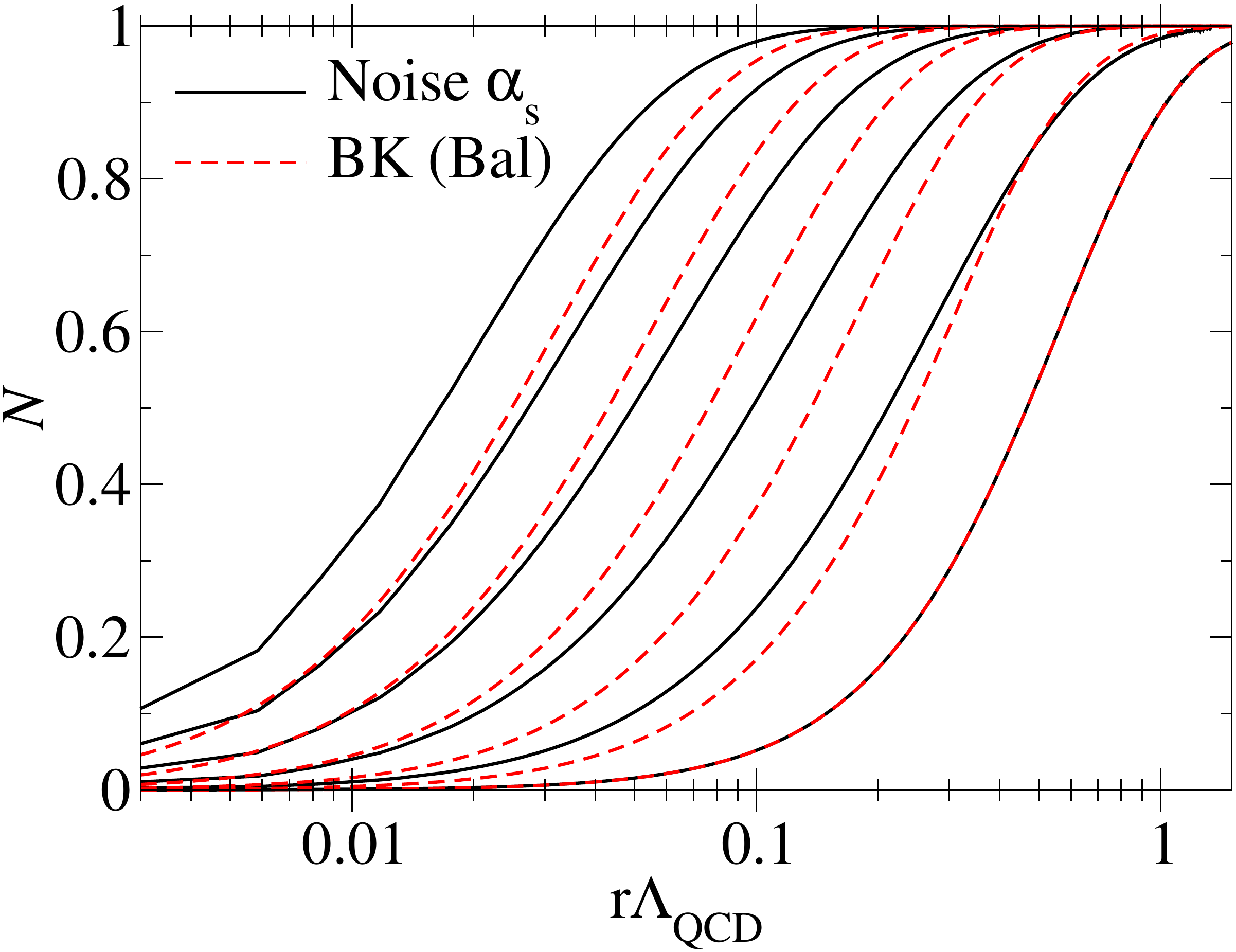}}
\caption{\label{fig:jvsbk}
Evolution of the scattering amplitude with the JIMWLK equation 
where the scale of the running coupling is set in the noise term,
\eq\nr{eq:newcorr}, compared to BK evolution with the
``Balitsky'' running coupling prescription, \eq\nr{eq:bal}.
The evolution is started from the same initial condition and the 
lines show the amplitude at intervals of 2 units in rapidity. 
}
\end{figure}

\begin{figure}[t]
\centerline{\includegraphics[width=0.45\textwidth]{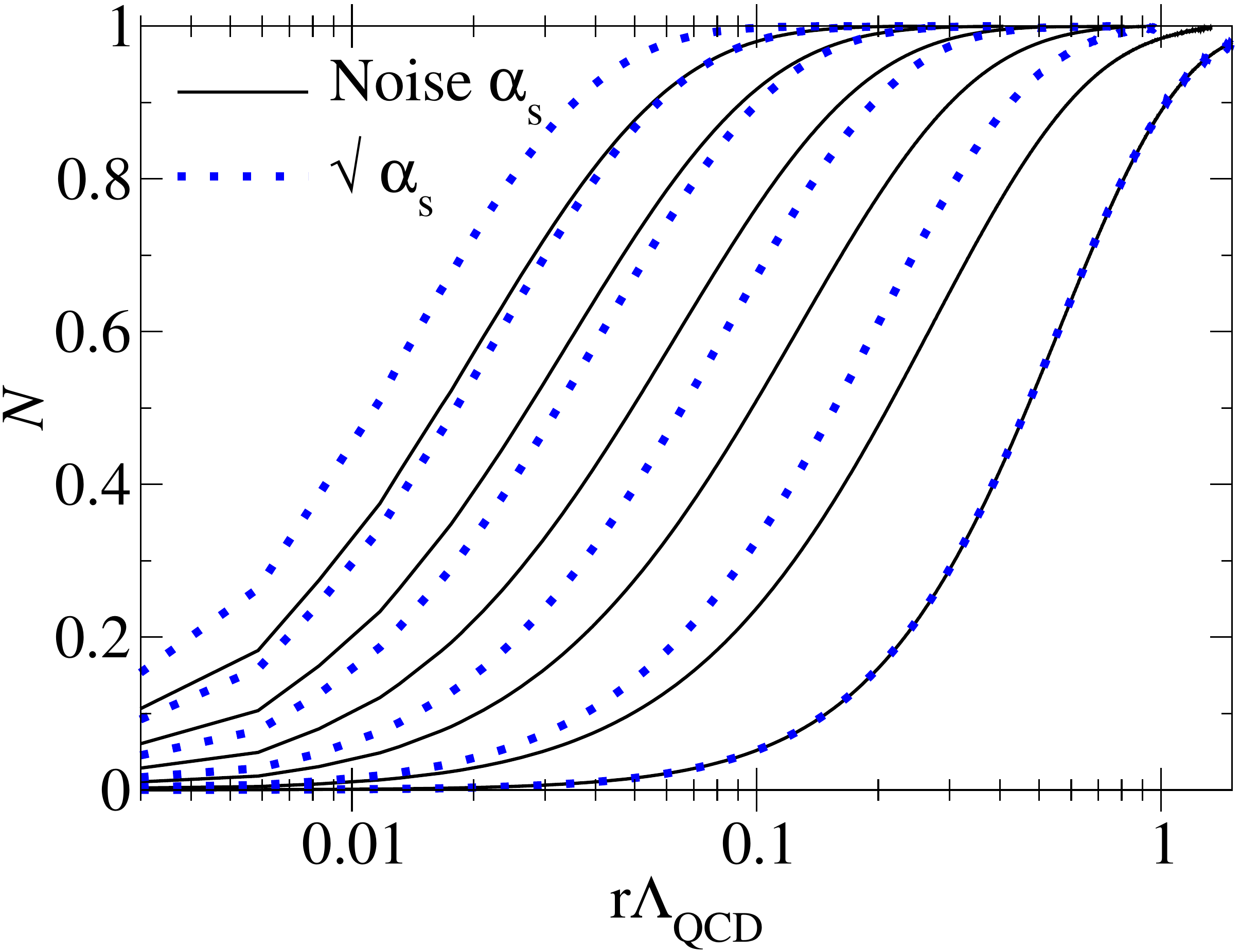}}
\caption{\label{fig:randvssqrt}
Evolution of the scattering amplitude $N \equiv 1-\langle\hat{D}_{\rt}\rangle$
with the  rcJIMWLK equation proposed in this paper 
(same data as in \fig\ref{fig:jvsbk}),
 compared to the ``square root'' prescription \nr{eq:sqrtas}
used in \cite{Lappi:2011ju,Dumitru:2011vk}. The evolution is started
from the same initial condition and the lines show the amplitude at 
intervals of 4 units in rapidity.
}
\end{figure}

\begin{figure}[t]
\centerline{\includegraphics[width=0.45\textwidth]{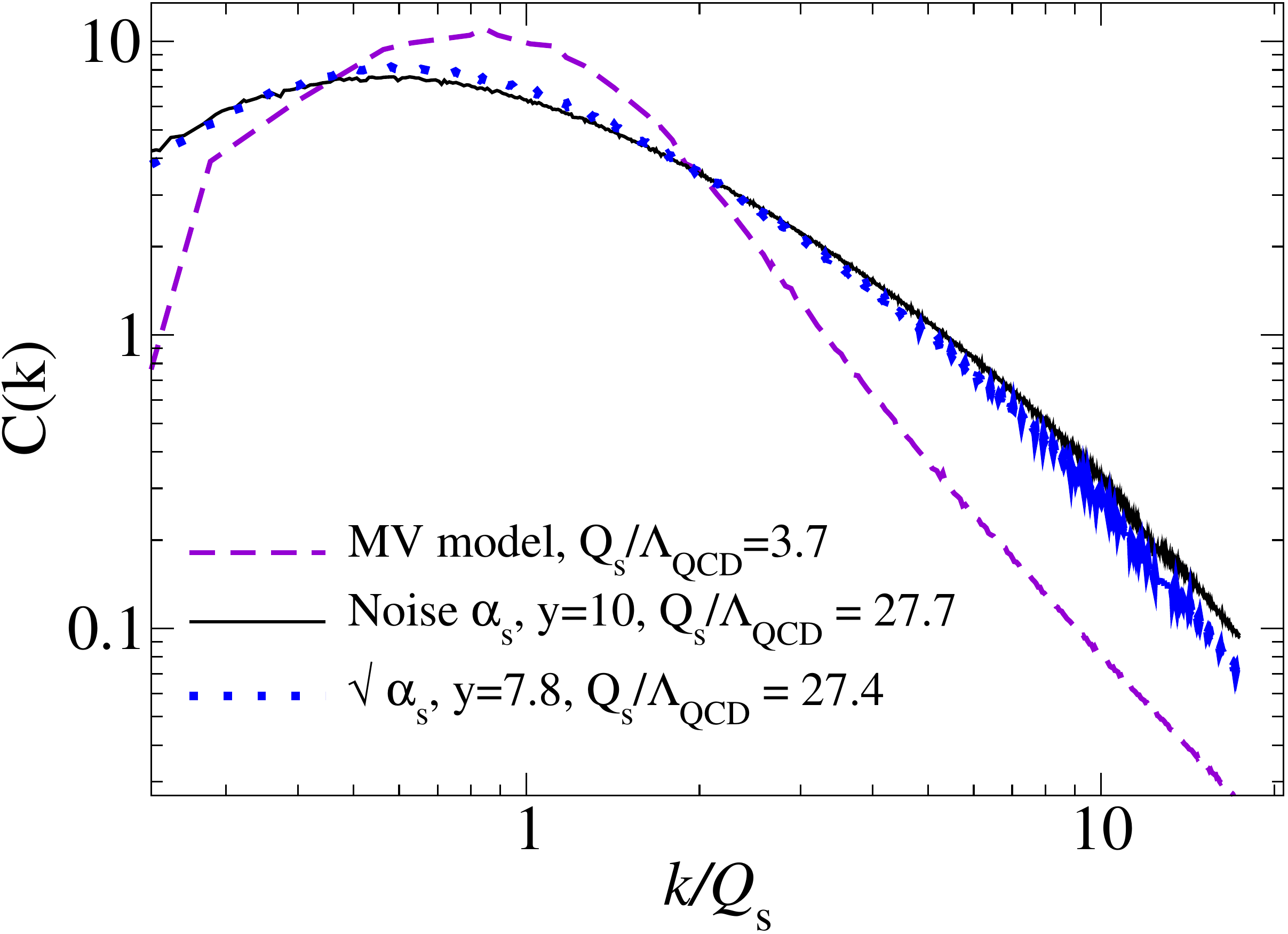}}
\caption{\label{fig:kspace}
The momentum space correlator, \nr{eq:kspace}, from  
the rcJIMWLK equation proposed in this paper compared to 
the ``square root'' prescription (with the same parameters 
as used in \fig\ref{fig:randvssqrt}). The correlator is shown for the 
initial condition, for which we use the MV model, 
and for $y=10$ and $y=7.8$ units of evolution for the
two running coupling versions; these correspond to approximately the
same value of $\qs$. Note that the end of the curves at large $\ktt$
corresponds to the lattice ultraviolet cutoffs, and the shape is 
sensitive to lattice effects well below that.
}
\end{figure}

\begin{figure}[t]
\centerline{\includegraphics[width=0.45\textwidth]{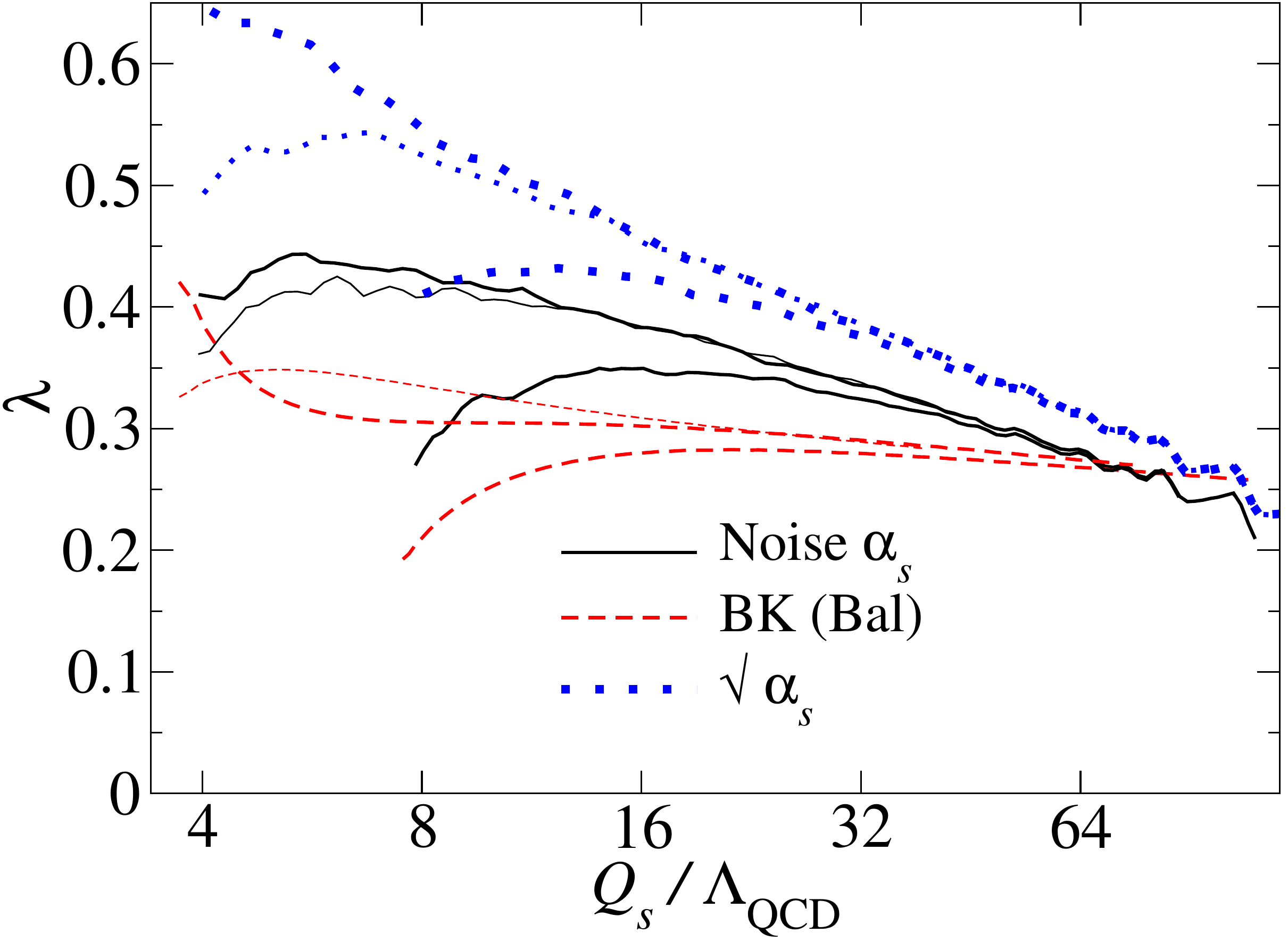}}
\caption{\label{fig:lambda}
Evolution speed of the saturation scale as a function of
$\qs/\lqcd$. Shown are the noise term (solid line), square root (dotted line)
running coupling JIMWLK equations and the Balitsky prescription BK equation 
(dashed line), starting from the MV model at two different initial
values of $\qs$. The thin lines have the freezing of the coupling
implmented more smoothly, with the value $c=1.5$ instead of $c=0.2$ as
used elsewhere in this paper.
}
\end{figure}

The numerical algorithm for solving the JIMWLK equation is discussed
in detail in \cite{Rummukainen:2003ns}, and the modifications required
to adapt it to our version of the JIMWLK equation,
\eqs\nr{eq:jtimestepsymmeta} and \nr{eq:newnoisecorr}, are straightforward.
We construct the initial condition Wilson line configurations
as described in \re\cite{Lappi:2007ku}, using a $1024^2$ transverse 
lattice, with the MV model parameter $g^2\mu L=31$ and the longitudinal extent in 
the MV model discretized in $N_y=100$ steps, leading to an initial 
fundamental representation saturation scale $\qs L \approx 22$, 
where $L$ is the size of the lattice in physical units.
The infrared behavior of the coupling is regularized as
\begin{equation}\label{eq:kspaceas}
  \as(\kt) = \frac{4\pi}{
\beta \ln\left\{ \left[
       \left(\frac{\mu_0^2}{\lqcd^2}\right)^{\frac{1}{c}}
      +\left( \frac{\kt^2}{\lqcd^2} \right)^{\frac{1}{c}} \right]^{c}
\right\} 
},
\end{equation}
with $\beta = 11-2\nf/3$, where we take $\nf=3$.
The  parameter $c=0.2$ regulates the sharpness of the freezing of the coupling, which 
happens at $\mu_0 L=15$. The value of the QCD scale is taken as
$\lqcd L = 6$, leading to a value $\alpha_0=0.7619$ for the coupling in the infrared.
In the more extensive study in \re\cite{Lappi:2011ju} the evolution speed was found
to be very insensitive to the lattice parameters and to $\alpha_0$ 
while naturally dependent on the physical ratio of scales $\qs/\lqcd$.

When the coupling is needed as a function of the transverse coordinate, 
as in the BK equation or the $\sqrt{\as}$-prescription for JIMWKL that 
we study as a comparison, we use 
\begin{equation}\label{eq:rspaceas}
  \as(\rt) = \frac{4\pi}{
\beta \ln\left\{\left[
       \left(\frac{\mu_0^2}{\lqcd^2}\right)^{\frac{1}{c}}
      +\left(\frac{4e^{-2 \gamma_\mathrm{E}}}{\rt^2\lqcd^2}\right)^{\frac{1}{c}} \right]^{c}
\right\}
}
\end{equation}
Note the constant factor $4e^{-2\gamma_\mathrm{E}}\approx 1.26$ in the identification 
$\ktt^2 \sim 4e^{-2\gamma_{\mathrm{E}}} /\rtt^2$, which is taken from the explicit
Fourier-transform of the kernel in \res\cite{Gardi:2006rp,Kovchegov:2006vj}.
In Appendix~\ref{app:numissues} we show a comparison between the coordinate 
and momentum  versions of the running coupling, verifying that this is indeed
the correct correspondence.
Phenomenological studies typically~\cite{Albacete:2009fh,Albacete:2010sy}
find that the coordinate space scale needs to be taken as larger,
$\sim 20/\rtt^2$ for $\lqcd \gtrsim 200$MeV, unless
some other corrections are applied that slow down the evolution,
such as in \re\cite{Kuokkanen:2011je}.
Thus, in spite of the slowness of the rcBK evolution with the
Balitsky prescription compared to fixed coupling, the data seems to prefer
a slower evolution still than it would predict.

Figure \ref{fig:jvsbk} compares the evolution of the dipole using
the rcJIMWLK equation  (\eqs\nr{eq:jtimestepsymmeta} and \nr{eq:newnoisecorr}
with~\nr{eq:kspaceas} for the coupling) with the Balitsky prescription 
rcBK equation \nr{eq:bal} with the coupling \nr{eq:rspaceas} 
over 20 units in rapidity. We see that the Balitsky prescription leads
still to a slower evolution than our proposal,
in spite of the similarity in parametrically different parts
of the phase space discussed above.
Discretization effects that also affect the comparison 
are discussed in Appendix~\ref{app:numissues}.
The shape of the correlator as a function of $\rtt$ is steeper in BK than 
in JIMWLK, for which we do not see a compelling explanation.

Figure \ref{fig:randvssqrt} compares the rcJIMWLK evolution using our proposal
to the ``square root'' coupling~\nr{eq:sqrtas} used in previous rcJIMWLK
calculations~\cite{Lappi:2011ju,Dumitru:2011vk}. 
One can see that the the evolution speed is slowed down significantly
from the ''square root'' prescription''.
While the shape of the front seems somewhat different
there is no clear sign of a different power law in the 
momentum space correlator
\begin{equation}\label{eq:kspace}
C(\kt) = \kt^2 \int \ud^2\rt e^{i\kt\cdot \rt }\left\langle \hat{D} \left(\rtt\right) \right\rangle.
\end{equation}
This is shown in \fig\ref{fig:kspace} as a function of the 
scaling variable $\ktt/\qs$ for both coupling constant prescriptions 
at a similar value of $\qs$ (which, due to the different evolution speed,
correspond to  slightly different values of rapidity). More detailed
statements on the geometric scaling behavior at large $\ktt$ would
require significantly larger lattices to push out the $\sim 1/a$
ultraviolet cutoff further.

Figure
\ref{fig:lambda} further illustrates the difference in evolution
speeds between the prescriptions. The evolution speed is defined as
\begin{equation}
\lambda \equiv \frac{\ud \ln \qs^2}{\ud y},
\end{equation}
with the saturation scale $\qs$ solved from  the condition
\begin{equation}
\left\langle \hat{D} \left(\rtt^2=2/\qs^2\right) \right\rangle = e^{-1/2}.
\end{equation}
In addition to the initial condition used in \fig\ref{fig:randvssqrt},
evolved for 20 units in $y$, \fig\ref{fig:lambda} also shows
the result of a calculation started with an initial $\qs$ twice as large.
The evolution speed shows the typical behavior also seen with the BK equation 
(see e.g.~\cite{Albacete:2007sm,Kuokkanen:2011je}), with a slower
evolution in the initial transient region when the dipole cross section
changes from the initial condition to the pseudoscaling form, followed
by a universal $1/\sqrt{y}$-dependence characteristic of rcBK evolution.
For comparison \fig\ref{fig:lambda}
also shows the evolution speed calculated with a smoother freezing
of the coupling in the infrared, implemented by choosing $c=1.5$ in 
\eqs\nr{eq:kspaceas} and~\nr{eq:rspaceas}. The dependence on the freezing 
prescription, which is not present for larger initial 
$\qs/\lqcd$ (not shown on the figure) is, interestingly, different
for the BK and JIMWLK calculations.
This initial behavior of course also
depends on the functional form of the initial condition.
At large rapidities, the evolution speed in JIMWLK drops below that in BK due 
to the fact that $\qs$ grows close to the lattice ultraviolet cutoff;
see the discussion in Appendix~\ref{app:numissues}.

We remind the reader that the typical phenomenologically favored initial saturation scale 
for the evolution is $\qs\sim 1\gev$ and the evolution speed resulting from fits
to HERA data 
(e.g.~\cite{Golec-Biernat:1998js,Iancu:2003ge,Soyez:2007kg,Albacete:2009fh,Albacete:2010sy,Kuokkanen:2011je} )
is $\lambda \sim 0.2\dots0.3$. The ``square root'' coupling
could be made to fit experimental data by adjusting the coefficient under the 
logarithm in the running coupling as $\ln [4C^2/(\rt^2\lqcd^2)]$ effectively 
taking $\lqcd$ as a fit parameter.
Doing so would, however, lead to an unnaturally large value for $C$, i.e. 
small value for the effective $\lqcd$. With our rcJIMWLK prescription, 
as with the Balitsky rcBK, this problem is much less severe, although still present
to some degree.

\section{Conclusions}
\label{sec:conc}

Until now running coupling solutions of the JIMWLK equation have 
been limited by the belief that since the Langevin formulation is written in 
terms of the square root of the JIMWLK kernel, one is limited to considering
only prescriptions that modify this square root. In particular, since
the square root only depends on the size of the daughter dipole, it has been 
impossible to include a dependence on the ``parent'' dipole size in the
coupling. We have in this paper pointed out that there is in fact another scale
also in the Langevin formulation that the running coupling can depend on, 
namely the scale in the random noise term.
We have  proposed to use this scale as the
argument of the running coupling, with the Langevin timestep of the
JIMWLK equation given by \eqs\nr{eq:jtimestepsymmeta} and \nr{eq:newnoisecorr}.
 Choosing this as the scale of
the coupling has several advantages. Firstly it corresponds naturally 
to the general argument that the scale of the coupling should be the momentum
of the emitted gluon. Secondly  we have shown analytically how
it leads to the typical scale of the coupling being taken as the 
smallest one of the three relevant dipole sizes (the ``parent'' and 
two ``daughter'' dipoles) present in a dipole splitting. This is 
very similar to the ``Balitsky'' prescription for the running coupling 
in the BK equation.
We have shown numerically that the evolution speed of the saturation scale
in our proposed running coupling JIMLWK equation is slower than the 
``square root'' coupling, although not fully consistent with the 
Balitsky prescription for BK, or with
fits of the data~\cite{Albacete:2009fh,Albacete:2010sy}.

\section*{Acknowledgements}
T.L. thanks F.~Gelis for discussions and IPhT, CEA/Saclay 
(URA du CNRS) for hospitality during the early stage of this work.
H.M. is supported by the Graduate School of Particle and Nuclear Physics.
This work has been supported by the Academy of Finland, project 
133005, and by computing resources from
CSC -- IT Center for Science in Espoo, Finland. 

\appendix

 \section{Lattice effects}\label{app:numissues}

\begin{figure}[t]
\centerline{\includegraphics[width=0.45\textwidth]{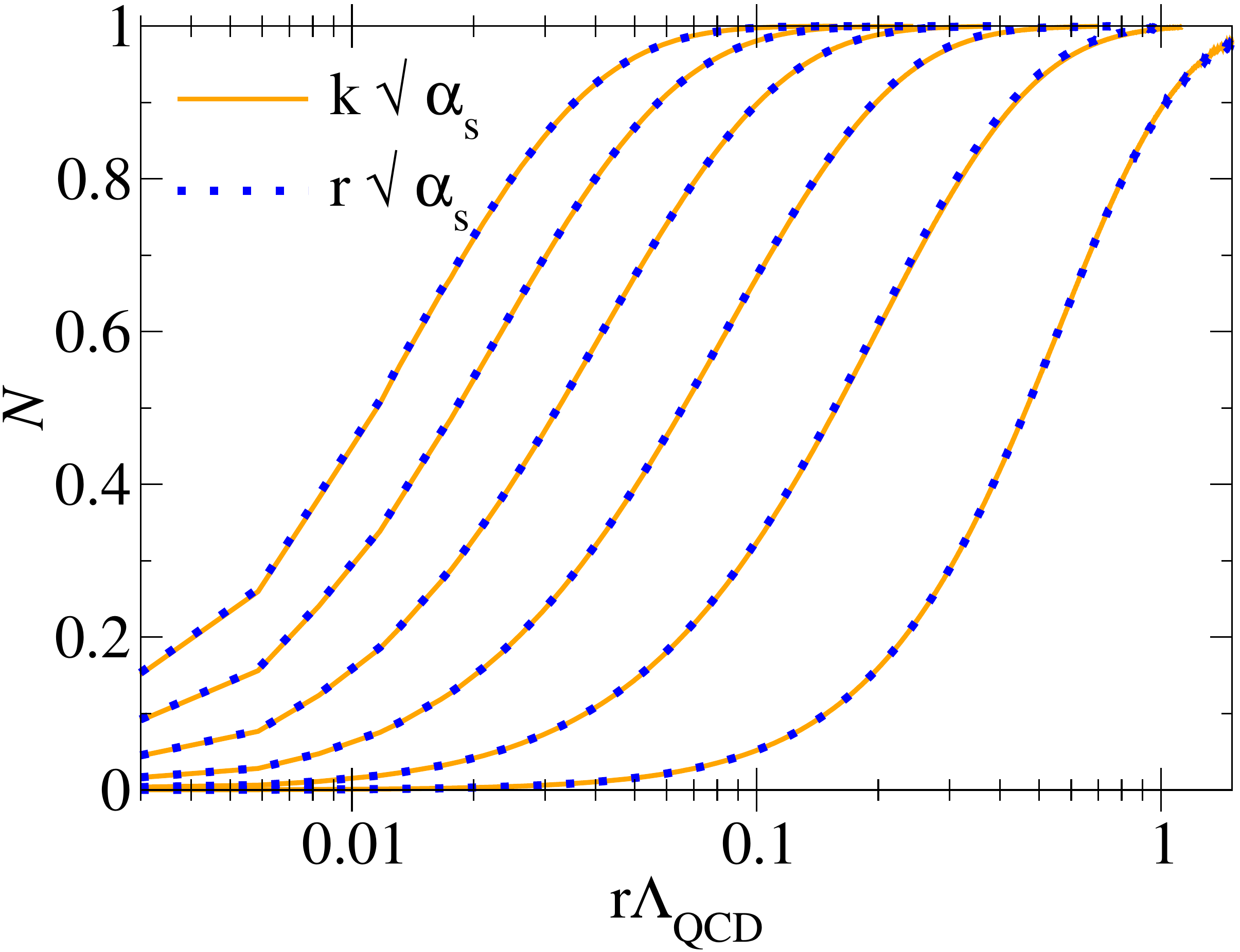}}
\caption{\label{fig:momsp}
Evolution of the scattering amplitude $N \equiv 1-\langle\hat{D}_{\rt}\rangle$
with the ``square root''  coupling rcJIMWLK equation, with coupling 
evaluated at the scale $\ktt^2$ in the momentum space kernel, and
at the scale $4e^{-2\gamma_\mathrm{E}}/\rtt^2$ in the position space kernel.
}
\end{figure}

\begin{figure}[t]
\centerline{\includegraphics[width=0.45\textwidth]{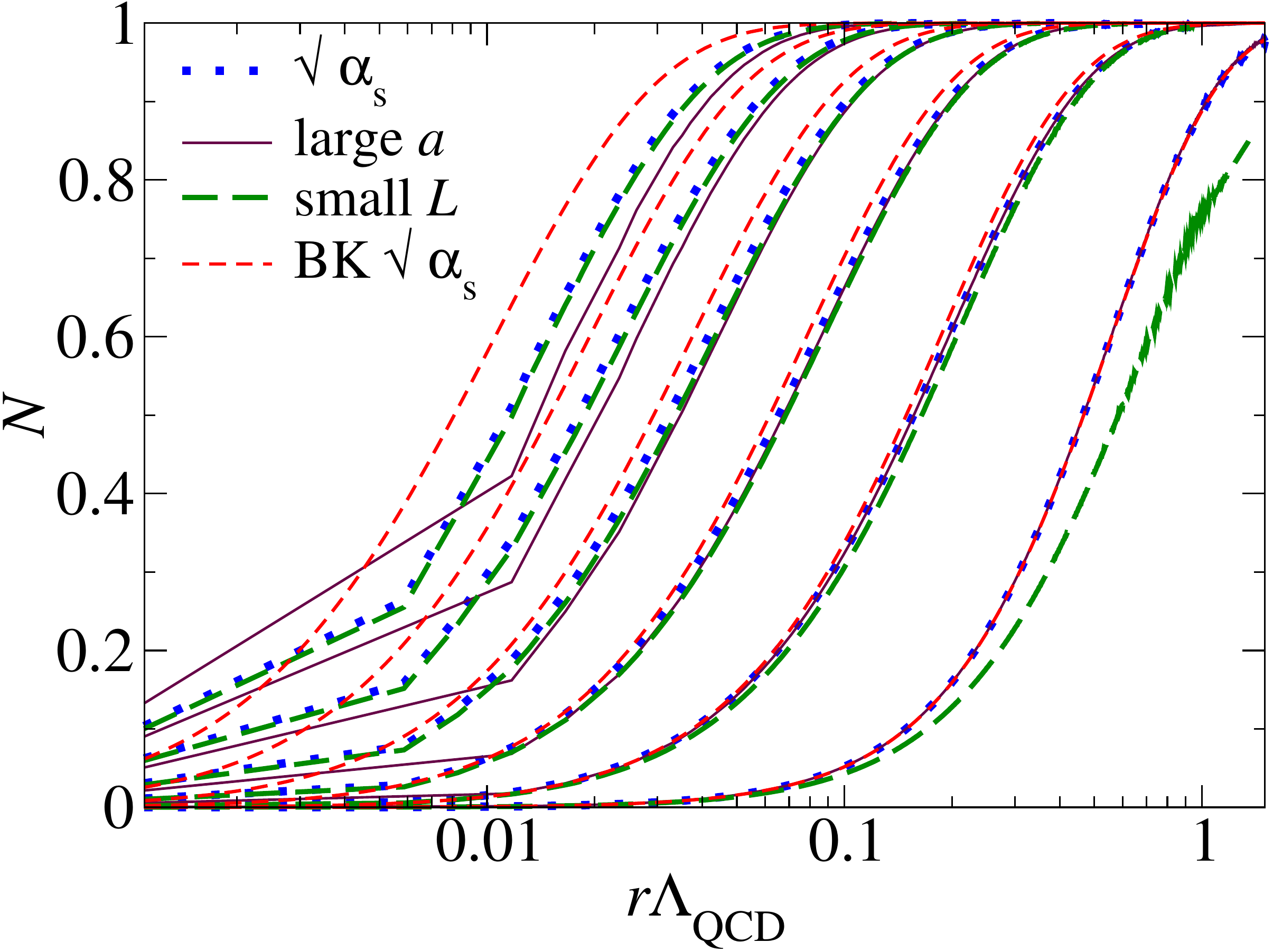}}
\caption{\label{fig:sqrtcomparison}
Evolution of the scattering amplitude $N \equiv 1-\langle\hat{D}_{\rt}\rangle$
with the ``square root''  rcJIMWLK equation (dotted line)  and with the BK equation 
(thinner dashed line)
using the same ``square root'' running coupling. The initial condition is the same as
in \fig\ref{fig:randvssqrt} and the amplitude is shown every 4 units in $y$.
Also shown are JIMWLK  simulations on a smaller $512^2$-lattice with the physical lattice 
size reduced (``small $L$'', thick dashed line) or lattice spacing 
increased (``large $a$'', thin solid line)  by a factor $2$.
}
\end{figure}

\begin{figure}[t]
\centerline{\includegraphics[width=0.45\textwidth]{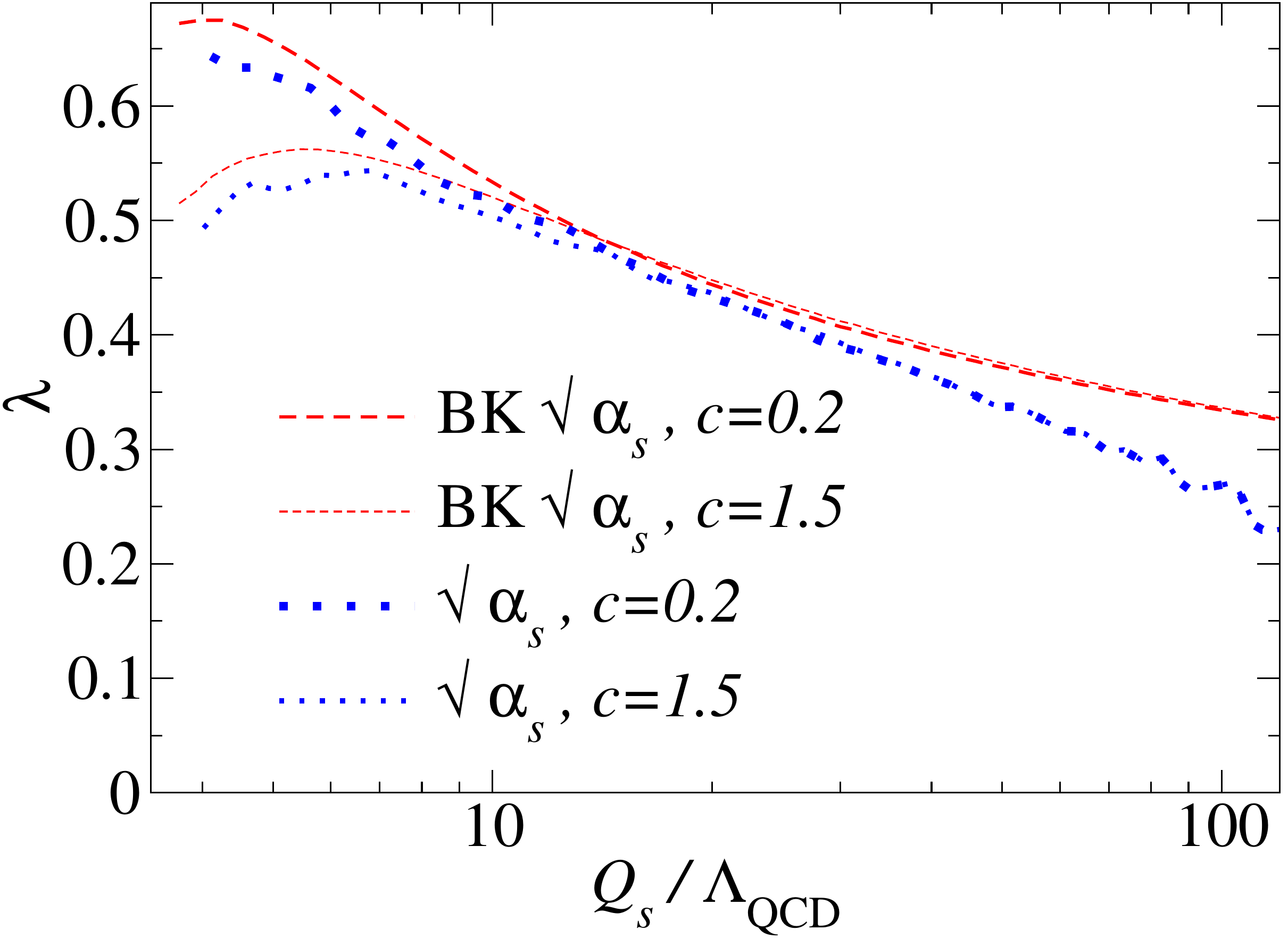}}
\caption{\label{fig:lambdaccomp}
Evolution speed of the saturation scale $\lambda$ with the ``square root'' coupling.
Shown are the results from the BK and JIMWLK simulations with the same kernel, 
using infrared regularizations of the running coupling with diferent
values of the parameter $c$.
}
\end{figure}

As test of the equivalence of the scales $\ktt^2$ in the momentum space coupling and
the scale $ 4e^{-2 \gamma_{\mathrm{E}}} /\rtt^2$ in coordinate space
we show in \fig\ref{fig:momsp} a comparison between a solution of the rcJIMWLK 
equation using the two couplings. This can be done strainghtforwardly when one uses
the ``square root'' coupling, where one can easily multply either the
kernel $\Kt_\xt$ or its Fourier-transform with $\sqrt{\as}$. Thus the curves in 
\fig\ref{fig:momsp} correspond to evolution using the kernels
\begin{equation}
\sqrt{\as(\rt)}\Kt_\rt,
\end{equation}
labeled $r \sqrt{\as}$, and
\begin{equation}
i(2\pi)^2 \int \frac{\ud^2\kt}{(2\pi)^2}
\sqrt{\as(\kt)}\frac{e^{-i \kt\cdot \xt}}{\kt^2},
\end{equation}
labeled $k \sqrt{\as}$, with $\as(\xt)$ and $\as(\kt)$ given by \eqs\nr{eq:kspaceas}
and~\nr{eq:rspaceas}. They are seen to agree remarkably well, 
justifying our use of the constant $ 4e^{-2\gamma_{\mathrm{E}}}$ 
in comparing the momentum and position space formulations.

The BK equation is solved on a logarithmic grid in $\rtt$, with the values
of the scattering amplitude between the grid points obtained from interpolation.
This allows one in practice to study as small or large $\rtt$-values
as desired. In the case of JIMWLK, on the other hand, one works on 
a fixed linear lattice with cutoffs $\sim 1/L$ in the infrared  and 
$\sim 1/a$  in the ultraviolet. One expects large lattice effects
when $\qs \sim 1/L$ and when $\qs \sim 1/a$, and since $\qs$ changes exponentially
during the evolution, this restricts the accessible rapidity range.

To illustrate the numerical uncertainties in our calculation
we compare in \fig\ref{fig:sqrtcomparison} the rcJIMWLK calculation
with a square root running coupling, shown also in \fig\ref{fig:randvssqrt},
to the BK equation with the same running coupling prescription. One can 
see that the solutions agree very well for smaller rapidities, but start to deviate
more later in the evolution. Also shown is a rcJIMWLK calculation with the
same initial $g^2\mu/\lqcd$ (approximately the same $\qs/\lqcd)$
performed on a smaller $512^2$-lattice in such a way that either the 
IR $\sim 1/L$ or UV $\sim 1/a$ cutoff is different. We see that for the initial condition 
$\qs$ is so small that changing the IR cutoff starts to have an effect. More importantly,
late in the evolution the solution becomes sensitive to the UV cutoff; an extrapolation
to $a \to 0 $  would bring the JIMWLK result closer to BK. 

A further comparison of the evolution speed in  the BK and rcJIMWLK codes is shown
in \fig\ref{fig:lambdaccomp}. It can be seen that in BK the evolution is slightly
faster, and increadingly so at large $\qs$, when the lattice ultraviolet cutoff
in the JIMWLK code slows down the evolution. In a  JIMWLK simulation closer
to the continuum limit the difference would be smaller. The figure also demostrates
the dependence on the parameter $c$ controlling the smoothness of the 
infrared regularization of the coupling. As was seen previously 
in \fig\ref{fig:lambda}, the evolution speed is sensitive to the 
regularization for small $\qs/\lqcd$.

In addition, the BK and JIMWLK equations are not equivalent even in the continuum, 
because of the mean field approximation.
 Making a precise statement
about this difference would require a more systematical extrapolation 
of the lattice calculation  to the continuum and infinite volume limits than
is done here. For fixed coupling
the difference in the evolution speed between the BK and JIMWLK equations
 has been studied in \re\cite{Kovchegov:2008mk}, where JIMWLK evolution
was found to be slower by few per cent, while an even smaller difference was 
seen for running coupling in \re\cite{kuokkanen}.

\bibliography{spires}

\ifx\mcitethebibliography\mciteundefinedmacro
\PackageError{JHEP-2modM.bst}{mciteplus.sty has not been loaded}
{This bibstyle requires the use of the mciteplus package.}\fi
\providecommand{\href}[2]{#2}
\begingroup\begin{mcitethebibliography}{10}

\bibitem{Gelis:2010nm}
F.~Gelis, E.~Iancu, J.~Jalilian-Marian and R.~Venugopalan
  \href{http://dx.doi.org/10.1146/annurev.nucl.010909.083629}{{\em Ann. Rev.
  Nucl. Part. Sci.} {\bf 60} (2010) 463}
  [\href{http://arXiv.org/abs/1002.0333}{{\tt arXiv:1002.0333 [hep-ph]}}]\relax
\mciteBstWouldAddEndPuncttrue
\mciteSetBstMidEndSepPunct{\mcitedefaultmidpunct}
{\mcitedefaultendpunct}{\mcitedefaultseppunct}\relax
\EndOfBibitem
\bibitem{Lappi:2010ek}
T.~Lappi \href{http://dx.doi.org/10.1142/S0218301311017302}{{\em Int. J. Mod.
  Phys.} {\bf E20} (2011) 1} [\href{http://arXiv.org/abs/1003.1852}{{\tt
  arXiv:1003.1852 [hep-ph]}}]\relax
\mciteBstWouldAddEndPuncttrue
\mciteSetBstMidEndSepPunct{\mcitedefaultmidpunct}
{\mcitedefaultendpunct}{\mcitedefaultseppunct}\relax
\EndOfBibitem
\bibitem{Jalilian-Marian:1997xn}
J.~Jalilian-Marian, A.~Kovner, L.~D. McLerran and H.~Weigert
  \href{http://dx.doi.org/10.1103/PhysRevD.55.5414}{{\em Phys. Rev.} {\bf D55}
  (1997) 5414} [\href{http://arXiv.org/abs/hep-ph/9606337}{{\tt
  arXiv:hep-ph/9606337}}]\relax
\mciteBstWouldAddEndPuncttrue
\mciteSetBstMidEndSepPunct{\mcitedefaultmidpunct}
{\mcitedefaultendpunct}{\mcitedefaultseppunct}\relax
\EndOfBibitem
\bibitem{Jalilian-Marian:1997jx}
J.~Jalilian-Marian, A.~Kovner, A.~Leonidov and H.~Weigert
  \href{http://dx.doi.org/10.1016/S0550-3213(97)00440-9}{{\em Nucl. Phys.} {\bf
  B504} (1997) 415} [\href{http://arXiv.org/abs/hep-ph/9701284}{{\tt
  arXiv:hep-ph/9701284}}]\relax
\mciteBstWouldAddEndPuncttrue
\mciteSetBstMidEndSepPunct{\mcitedefaultmidpunct}
{\mcitedefaultendpunct}{\mcitedefaultseppunct}\relax
\EndOfBibitem
\bibitem{Jalilian-Marian:1997gr}
J.~Jalilian-Marian, A.~Kovner, A.~Leonidov and H.~Weigert
  \href{http://dx.doi.org/10.1103/PhysRevD.59.014014}{{\em Phys. Rev.} {\bf
  D59} (1999) 014014} [\href{http://arXiv.org/abs/hep-ph/9706377}{{\tt
  arXiv:hep-ph/9706377}}]\relax
\mciteBstWouldAddEndPuncttrue
\mciteSetBstMidEndSepPunct{\mcitedefaultmidpunct}
{\mcitedefaultendpunct}{\mcitedefaultseppunct}\relax
\EndOfBibitem
\bibitem{Jalilian-Marian:1997dw}
J.~Jalilian-Marian, A.~Kovner and H.~Weigert
  \href{http://dx.doi.org/10.1103/PhysRevD.59.014015}{{\em Phys. Rev.} {\bf
  D59} (1999) 014015} [\href{http://arXiv.org/abs/hep-ph/9709432}{{\tt
  arXiv:hep-ph/9709432}}]\relax
\mciteBstWouldAddEndPuncttrue
\mciteSetBstMidEndSepPunct{\mcitedefaultmidpunct}
{\mcitedefaultendpunct}{\mcitedefaultseppunct}\relax
\EndOfBibitem
\bibitem{JalilianMarian:1998cb}
J.~Jalilian-Marian, A.~Kovner, A.~Leonidov and H.~Weigert
  \href{http://dx.doi.org/10.1103/PhysRevD.59.034007}{{\em Phys. Rev.} {\bf
  D59} (1999) 034007} [\href{http://arXiv.org/abs/hep-ph/9807462}{{\tt
  arXiv:hep-ph/9807462}}]\relax
\mciteBstWouldAddEndPuncttrue
\mciteSetBstMidEndSepPunct{\mcitedefaultmidpunct}
{\mcitedefaultendpunct}{\mcitedefaultseppunct}\relax
\EndOfBibitem
\bibitem{Iancu:2000hn}
E.~Iancu, A.~Leonidov and L.~D. McLerran
  \href{http://dx.doi.org/10.1016/S0375-9474(01)00642-X}{{\em Nucl. Phys.} {\bf
  A692} (2001) 583} [\href{http://arXiv.org/abs/hep-ph/0011241}{{\tt
  arXiv:hep-ph/0011241}}]\relax
\mciteBstWouldAddEndPuncttrue
\mciteSetBstMidEndSepPunct{\mcitedefaultmidpunct}
{\mcitedefaultendpunct}{\mcitedefaultseppunct}\relax
\EndOfBibitem
\bibitem{Iancu:2001md}
E.~Iancu and L.~D. McLerran
  \href{http://dx.doi.org/10.1016/S0370-2693(01)00526-3}{{\em Phys. Lett.} {\bf
  B510} (2001) 145} [\href{http://arXiv.org/abs/hep-ph/0103032}{{\tt
  arXiv:hep-ph/0103032}}]\relax
\mciteBstWouldAddEndPuncttrue
\mciteSetBstMidEndSepPunct{\mcitedefaultmidpunct}
{\mcitedefaultendpunct}{\mcitedefaultseppunct}\relax
\EndOfBibitem
\bibitem{Ferreiro:2001qy}
E.~Ferreiro, E.~Iancu, A.~Leonidov and L.~McLerran
  \href{http://dx.doi.org/10.1016/S0375-9474(01)01329-X}{{\em Nucl. Phys.} {\bf
  A703} (2002) 489} [\href{http://arXiv.org/abs/hep-ph/0109115}{{\tt
  arXiv:hep-ph/0109115}}]\relax
\mciteBstWouldAddEndPuncttrue
\mciteSetBstMidEndSepPunct{\mcitedefaultmidpunct}
{\mcitedefaultendpunct}{\mcitedefaultseppunct}\relax
\EndOfBibitem
\bibitem{Iancu:2001ad}
E.~Iancu, A.~Leonidov and L.~D. McLerran
  \href{http://dx.doi.org/10.1016/S0370-2693(01)00524-X}{{\em Phys. Lett.} {\bf
  B510} (2001) 133} [\href{http://arXiv.org/abs/hep-ph/0102009}{{\tt
  arXiv:hep-ph/0102009}}]\relax
\mciteBstWouldAddEndPuncttrue
\mciteSetBstMidEndSepPunct{\mcitedefaultmidpunct}
{\mcitedefaultendpunct}{\mcitedefaultseppunct}\relax
\EndOfBibitem
\bibitem{Weigert:2000gi}
H.~Weigert \href{http://dx.doi.org/10.1016/S0375-9474(01)01668-2}{{\em Nucl.
  Phys.} {\bf A703} (2002) 823}
  [\href{http://arXiv.org/abs/hep-ph/0004044}{{\tt arXiv:hep-ph/0004044
  [hep-ph]}}]\relax
\mciteBstWouldAddEndPuncttrue
\mciteSetBstMidEndSepPunct{\mcitedefaultmidpunct}
{\mcitedefaultendpunct}{\mcitedefaultseppunct}\relax
\EndOfBibitem
\bibitem{Mueller:2001uk}
A.~H. Mueller \href{http://dx.doi.org/10.1016/S0370-2693(01)01343-0}{{\em Phys.
  Lett.} {\bf B523} (2001) 243}
  [\href{http://arXiv.org/abs/hep-ph/0110169}{{\tt
  arXiv:hep-ph/0110169}}]\relax
\mciteBstWouldAddEndPuncttrue
\mciteSetBstMidEndSepPunct{\mcitedefaultmidpunct}
{\mcitedefaultendpunct}{\mcitedefaultseppunct}\relax
\EndOfBibitem
\bibitem{Balitsky:1995ub}
I.~Balitsky \href{http://dx.doi.org/10.1016/0550-3213(95)00638-9}{{\em Nucl.
  Phys.} {\bf B463} (1996) 99} [\href{http://arXiv.org/abs/hep-ph/9509348}{{\tt
  arXiv:hep-ph/9509348}}]\relax
\mciteBstWouldAddEndPuncttrue
\mciteSetBstMidEndSepPunct{\mcitedefaultmidpunct}
{\mcitedefaultendpunct}{\mcitedefaultseppunct}\relax
\EndOfBibitem
\bibitem{Kovchegov:1999yj}
Y.~V. Kovchegov \href{http://dx.doi.org/10.1103/PhysRevD.60.034008}{{\em Phys.
  Rev.} {\bf D60} (1999) 034008}
  [\href{http://arXiv.org/abs/hep-ph/9901281}{{\tt
  arXiv:hep-ph/9901281}}]\relax
\mciteBstWouldAddEndPuncttrue
\mciteSetBstMidEndSepPunct{\mcitedefaultmidpunct}
{\mcitedefaultendpunct}{\mcitedefaultseppunct}\relax
\EndOfBibitem
\bibitem{Kovchegov:1999ua}
Y.~V. Kovchegov \href{http://dx.doi.org/10.1103/PhysRevD.61.074018}{{\em Phys.
  Rev.} {\bf D61} (2000) 074018}
  [\href{http://arXiv.org/abs/hep-ph/9905214}{{\tt
  arXiv:hep-ph/9905214}}]\relax
\mciteBstWouldAddEndPuncttrue
\mciteSetBstMidEndSepPunct{\mcitedefaultmidpunct}
{\mcitedefaultendpunct}{\mcitedefaultseppunct}\relax
\EndOfBibitem
\bibitem{Marquet:2007vb}
C.~Marquet \href{http://dx.doi.org/10.1016/j.nuclphysa.2007.09.001}{{\em Nucl.
  Phys.} {\bf A796} (2007) 41} [\href{http://arXiv.org/abs/0708.0231}{{\tt
  arXiv:0708.0231 [hep-ph]}}]\relax
\mciteBstWouldAddEndPuncttrue
\mciteSetBstMidEndSepPunct{\mcitedefaultmidpunct}
{\mcitedefaultendpunct}{\mcitedefaultseppunct}\relax
\EndOfBibitem
\bibitem{Albacete:2010pg}
J.~L. Albacete and C.~Marquet
  \href{http://dx.doi.org/10.1103/PhysRevLett.105.162301}{{\em Phys. Rev.
  Lett.} {\bf 105} (2010) 162301} [\href{http://arXiv.org/abs/1005.4065}{{\tt
  arXiv:1005.4065 [hep-ph]}}]\relax
\mciteBstWouldAddEndPuncttrue
\mciteSetBstMidEndSepPunct{\mcitedefaultmidpunct}
{\mcitedefaultendpunct}{\mcitedefaultseppunct}\relax
\EndOfBibitem
\bibitem{Dumitru:2011vk}
A.~Dumitru, J.~Jalilian-Marian, T.~Lappi, B.~Schenke and R.~Venugopalan
  \href{http://dx.doi.org/10.1016/j.physletb.2011.11.002}{{\em Phys. Lett.}
  {\bf B706} (2011) 219} [\href{http://arXiv.org/abs/1108.4764}{{\tt
  arXiv:1108.4764 [hep-ph]}}]\relax
\mciteBstWouldAddEndPuncttrue
\mciteSetBstMidEndSepPunct{\mcitedefaultmidpunct}
{\mcitedefaultendpunct}{\mcitedefaultseppunct}\relax
\EndOfBibitem
\bibitem{Dominguez:2011wm}
F.~Dominguez, C.~Marquet, B.-W. Xiao and F.~Yuan
  \href{http://dx.doi.org/10.1103/PhysRevD.83.105005}{{\em Phys. Rev.} {\bf
  D83} (2011) 105005} [\href{http://arXiv.org/abs/1101.0715}{{\tt
  arXiv:1101.0715 [hep-ph]}}]\relax
\mciteBstWouldAddEndPuncttrue
\mciteSetBstMidEndSepPunct{\mcitedefaultmidpunct}
{\mcitedefaultendpunct}{\mcitedefaultseppunct}\relax
\EndOfBibitem
\bibitem{Lappi:2012nh}
T.~Lappi and H.~M{\"a}ntysaari \href{http://arXiv.org/abs/1209.2853}{{\tt
  arXiv:1209.2853 [hep-ph]}}\relax
\mciteBstWouldAddEndPuncttrue
\mciteSetBstMidEndSepPunct{\mcitedefaultmidpunct}
{\mcitedefaultendpunct}{\mcitedefaultseppunct}\relax
\EndOfBibitem
\bibitem{Dumitru:2008wn}
A.~Dumitru, F.~Gelis, L.~McLerran and R.~Venugopalan
  \href{http://dx.doi.org/10.1016/j.nuclphysa.2008.06.012}{{\em Nucl. Phys.}
  {\bf A810} (2008) 91} [\href{http://arXiv.org/abs/0804.3858}{{\tt
  arXiv:0804.3858 [hep-ph]}}]\relax
\mciteBstWouldAddEndPuncttrue
\mciteSetBstMidEndSepPunct{\mcitedefaultmidpunct}
{\mcitedefaultendpunct}{\mcitedefaultseppunct}\relax
\EndOfBibitem
\bibitem{Gelis:2008sz}
F.~Gelis, T.~Lappi and R.~Venugopalan
  \href{http://dx.doi.org/10.1103/PhysRevD.79.094017}{{\em Phys. Rev.} {\bf
  D79} (2008) 094017} [\href{http://arXiv.org/abs/0810.4829}{{\tt
  arXiv:0810.4829 [hep-ph]}}]\relax
\mciteBstWouldAddEndPuncttrue
\mciteSetBstMidEndSepPunct{\mcitedefaultmidpunct}
{\mcitedefaultendpunct}{\mcitedefaultseppunct}\relax
\EndOfBibitem
\bibitem{Dusling:2009ni}
K.~Dusling, F.~Gelis, T.~Lappi and R.~Venugopalan
  \href{http://dx.doi.org/10.1016/j.nuclphysa.2009.12.044}{{\em Nucl. Phys.}
  {\bf A836} (2010) 159} [\href{http://arXiv.org/abs/0911.2720}{{\tt
  arXiv:0911.2720 [hep-ph]}}]\relax
\mciteBstWouldAddEndPuncttrue
\mciteSetBstMidEndSepPunct{\mcitedefaultmidpunct}
{\mcitedefaultendpunct}{\mcitedefaultseppunct}\relax
\EndOfBibitem
\bibitem{Dumitru:2010iy}
A.~Dumitru, K.~Dusling, F.~Gelis, J.~Jalilian-Marian, T.~Lappi and
  R.~Venugopalan \href{http://dx.doi.org/10.1016/j.physletb.2011.01.024}{{\em
  Phys. Lett.} {\bf B697} (2011) 21}
  [\href{http://arXiv.org/abs/1009.5295}{{\tt arXiv:1009.5295 [hep-ph]}}]\relax
\mciteBstWouldAddEndPuncttrue
\mciteSetBstMidEndSepPunct{\mcitedefaultmidpunct}
{\mcitedefaultendpunct}{\mcitedefaultseppunct}\relax
\EndOfBibitem
\bibitem{Dusling:2012ig}
K.~Dusling and R.~Venugopalan
  \href{http://dx.doi.org/10.1103/PhysRevLett.108.262001}{{\em Phys. Rev.
  Lett.} {\bf 108} (2012) 262001} [\href{http://arXiv.org/abs/1201.2658}{{\tt
  arXiv:1201.2658 [hep-ph]}}]\relax
\mciteBstWouldAddEndPuncttrue
\mciteSetBstMidEndSepPunct{\mcitedefaultmidpunct}
{\mcitedefaultendpunct}{\mcitedefaultseppunct}\relax
\EndOfBibitem
\bibitem{Kovchegov:2006vj}
Y.~V. Kovchegov and H.~Weigert
  \href{http://dx.doi.org/10.1016/j.nuclphysa.2006.10.075}{{\em Nucl. Phys.}
  {\bf A784} (2007) 188} [\href{http://arXiv.org/abs/hep-ph/0609090}{{\tt
  arXiv:hep-ph/0609090}}]\relax
\mciteBstWouldAddEndPuncttrue
\mciteSetBstMidEndSepPunct{\mcitedefaultmidpunct}
{\mcitedefaultendpunct}{\mcitedefaultseppunct}\relax
\EndOfBibitem
\bibitem{Balitsky:2006wa}
I.~Balitsky \href{http://dx.doi.org/10.1103/PhysRevD.75.014001}{{\em Phys.
  Rev.} {\bf D75} (2007) 014001}
  [\href{http://arXiv.org/abs/hep-ph/0609105}{{\tt
  arXiv:hep-ph/0609105}}]\relax
\mciteBstWouldAddEndPuncttrue
\mciteSetBstMidEndSepPunct{\mcitedefaultmidpunct}
{\mcitedefaultendpunct}{\mcitedefaultseppunct}\relax
\EndOfBibitem
\bibitem{Albacete:2007yr}
J.~L. Albacete and Y.~V. Kovchegov
  \href{http://dx.doi.org/10.1103/PhysRevD.75.125021}{{\em Phys. Rev.} {\bf
  D75} (2007) 125021} [\href{http://arXiv.org/abs/0704.0612}{{\tt
  arXiv:0704.0612 [hep-ph]}}]\relax
\mciteBstWouldAddEndPuncttrue
\mciteSetBstMidEndSepPunct{\mcitedefaultmidpunct}
{\mcitedefaultendpunct}{\mcitedefaultseppunct}\relax
\EndOfBibitem
\bibitem{Albacete:2009fh}
J.~L. Albacete, N.~Armesto, J.~G. Milhano and C.~A. Salgado
  \href{http://dx.doi.org/10.1103/PhysRevD.80.034031}{{\em Phys. Rev.} {\bf
  D80} (2009) 034031} [\href{http://arXiv.org/abs/0902.1112}{{\tt
  arXiv:0902.1112 [hep-ph]}}]\relax
\mciteBstWouldAddEndPuncttrue
\mciteSetBstMidEndSepPunct{\mcitedefaultmidpunct}
{\mcitedefaultendpunct}{\mcitedefaultseppunct}\relax
\EndOfBibitem
\bibitem{Albacete:2010sy}
J.~L. Albacete, N.~Armesto, J.~G. Milhano, P.~Quiroga-Arias and C.~A. Salgado
  \href{http://dx.doi.org/10.1140/epjc/s10052-011-1705-3}{{\em Eur. Phys. J.}
  {\bf C71} (2011) 1705} [\href{http://arXiv.org/abs/1012.4408}{{\tt
  arXiv:1012.4408 [hep-ph]}}]\relax
\mciteBstWouldAddEndPuncttrue
\mciteSetBstMidEndSepPunct{\mcitedefaultmidpunct}
{\mcitedefaultendpunct}{\mcitedefaultseppunct}\relax
\EndOfBibitem
\bibitem{Albacete:2010bs}
J.~L. Albacete and C.~Marquet
  \href{http://dx.doi.org/10.1016/j.physletb.2010.02.073}{{\em Phys. Lett.}
  {\bf B687} (2010) 174} [\href{http://arXiv.org/abs/1001.1378}{{\tt
  arXiv:1001.1378 [hep-ph]}}]\relax
\mciteBstWouldAddEndPuncttrue
\mciteSetBstMidEndSepPunct{\mcitedefaultmidpunct}
{\mcitedefaultendpunct}{\mcitedefaultseppunct}\relax
\EndOfBibitem
\bibitem{Kuokkanen:2011je}
J.~Kuokkanen, K.~Rummukainen and H.~Weigert
  \href{http://dx.doi.org/10.1016/j.nuclphysa.2011.10.006}{{\em Nucl. Phys.}
  {\bf A875} (2012) 29} [\href{http://arXiv.org/abs/1108.1867}{{\tt
  arXiv:1108.1867 [hep-ph]}}]\relax
\mciteBstWouldAddEndPuncttrue
\mciteSetBstMidEndSepPunct{\mcitedefaultmidpunct}
{\mcitedefaultendpunct}{\mcitedefaultseppunct}\relax
\EndOfBibitem
\bibitem{Albacete:2012xq}
J.~L. Albacete, A.~Dumitru, H.~Fujii and Y.~Nara
  \href{http://arXiv.org/abs/1209.2001}{{\tt arXiv:1209.2001 [hep-ph]}}\relax
\mciteBstWouldAddEndPuncttrue
\mciteSetBstMidEndSepPunct{\mcitedefaultmidpunct}
{\mcitedefaultendpunct}{\mcitedefaultseppunct}\relax
\EndOfBibitem
\bibitem{Balitsky:2008zza}
I.~Balitsky and G.~A. Chirilli
  \href{http://dx.doi.org/10.1103/PhysRevD.77.014019}{{\em Phys.Rev.} {\bf D77}
  (2008) 014019} [\href{http://arXiv.org/abs/0710.4330}{{\tt arXiv:0710.4330
  [hep-ph]}}]\relax
\mciteBstWouldAddEndPuncttrue
\mciteSetBstMidEndSepPunct{\mcitedefaultmidpunct}
{\mcitedefaultendpunct}{\mcitedefaultseppunct}\relax
\EndOfBibitem
\bibitem{Avsar:2011ds}
E.~Avsar, A.~Stasto, D.~Triantafyllopoulos and D.~Zaslavsky
  \href{http://dx.doi.org/10.1007/JHEP10(2011)138}{{\em JHEP} {\bf 1110} (2011)
  138} [\href{http://arXiv.org/abs/1107.1252}{{\tt arXiv:1107.1252
  [hep-ph]}}]\relax
\mciteBstWouldAddEndPuncttrue
\mciteSetBstMidEndSepPunct{\mcitedefaultmidpunct}
{\mcitedefaultendpunct}{\mcitedefaultseppunct}\relax
\EndOfBibitem
\bibitem{Lappi:2011ju}
T.~Lappi \href{http://dx.doi.org/10.1016/j.physletb.2011.08.011}{{\em Phys.
  Lett.} {\bf B703} (2011) 325} [\href{http://arXiv.org/abs/1105.5511}{{\tt
  arXiv:1105.5511 [hep-ph]}}]\relax
\mciteBstWouldAddEndPuncttrue
\mciteSetBstMidEndSepPunct{\mcitedefaultmidpunct}
{\mcitedefaultendpunct}{\mcitedefaultseppunct}\relax
\EndOfBibitem
\bibitem{Kovner:2005jc}
A.~Kovner and M.~Lublinsky
  \href{http://dx.doi.org/10.1088/1126-6708/2005/03/001}{{\em JHEP} {\bf 0503}
  (2005) 001} [\href{http://arXiv.org/abs/hep-ph/0502071}{{\tt
  arXiv:hep-ph/0502071 [hep-ph]}}]\relax
\mciteBstWouldAddEndPuncttrue
\mciteSetBstMidEndSepPunct{\mcitedefaultmidpunct}
{\mcitedefaultendpunct}{\mcitedefaultseppunct}\relax
\EndOfBibitem
\bibitem{Kovner:2005en}
A.~Kovner and M.~Lublinsky
  \href{http://dx.doi.org/10.1103/PhysRevLett.94.181603}{{\em Phys. Rev. Lett.}
  {\bf 94} (2005) 181603} [\href{http://arXiv.org/abs/hep-ph/0502119}{{\tt
  arXiv:hep-ph/0502119 [hep-ph]}}]\relax
\mciteBstWouldAddEndPuncttrue
\mciteSetBstMidEndSepPunct{\mcitedefaultmidpunct}
{\mcitedefaultendpunct}{\mcitedefaultseppunct}\relax
\EndOfBibitem
\bibitem{Iancu:2011nj}
E.~Iancu and D.~Triantafyllopoulos
  \href{http://dx.doi.org/10.1007/JHEP04(2012)025}{{\em JHEP} {\bf 1204} (2012)
  025} [\href{http://arXiv.org/abs/1112.1104}{{\tt arXiv:1112.1104
  [hep-ph]}}]\relax
\mciteBstWouldAddEndPuncttrue
\mciteSetBstMidEndSepPunct{\mcitedefaultmidpunct}
{\mcitedefaultendpunct}{\mcitedefaultseppunct}\relax
\EndOfBibitem
\bibitem{Blaizot:2002xy}
J.-P. Blaizot, E.~Iancu and H.~Weigert
  \href{http://dx.doi.org/10.1016/S0375-9474(02)01299-X}{{\em Nucl. Phys.} {\bf
  A713} (2003) 441} [\href{http://arXiv.org/abs/hep-ph/0206279}{{\tt
  arXiv:hep-ph/0206279 [hep-ph]}}]\relax
\mciteBstWouldAddEndPuncttrue
\mciteSetBstMidEndSepPunct{\mcitedefaultmidpunct}
{\mcitedefaultendpunct}{\mcitedefaultseppunct}\relax
\EndOfBibitem
\bibitem{Rummukainen:2003ns}
K.~Rummukainen and H.~Weigert
  \href{http://dx.doi.org/10.1016/j.nuclphysa.2004.03.219}{{\em Nucl. Phys.}
  {\bf A739} (2004) 183} [\href{http://arXiv.org/abs/hep-ph/0309306}{{\tt
  arXiv:hep-ph/0309306 [hep-ph]}}]\relax
\mciteBstWouldAddEndPuncttrue
\mciteSetBstMidEndSepPunct{\mcitedefaultmidpunct}
{\mcitedefaultendpunct}{\mcitedefaultseppunct}\relax
\EndOfBibitem
\bibitem{Kovchegov:2008mk}
Y.~V. Kovchegov, J.~Kuokkanen, K.~Rummukainen and H.~Weigert
  \href{http://dx.doi.org/10.1016/j.nuclphysa.2009.03.006}{{\em Nucl. Phys.}
  {\bf A823} (2009) 47} [\href{http://arXiv.org/abs/0812.3238}{{\tt
  arXiv:0812.3238 [hep-ph]}}]\relax
\mciteBstWouldAddEndPuncttrue
\mciteSetBstMidEndSepPunct{\mcitedefaultmidpunct}
{\mcitedefaultendpunct}{\mcitedefaultseppunct}\relax
\EndOfBibitem
\bibitem{Marquet:2010cf}
C.~Marquet and H.~Weigert
  \href{http://dx.doi.org/10.1016/j.nuclphysa.2010.05.056}{{\em Nucl. Phys.}
  {\bf A843} (2010) 68} [\href{http://arXiv.org/abs/1003.0813}{{\tt
  arXiv:1003.0813 [hep-ph]}}]\relax
\mciteBstWouldAddEndPuncttrue
\mciteSetBstMidEndSepPunct{\mcitedefaultmidpunct}
{\mcitedefaultendpunct}{\mcitedefaultseppunct}\relax
\EndOfBibitem
\bibitem{Cooley:1965zz}
J.~W. Cooley and J.~W. Tukey
  \href{http://dx.doi.org/10.1090/S0025-5718-1965-0178586-1}{{\em Math.
  Comput.} {\bf 19} (1965) 297}\relax
\mciteBstWouldAddEndPuncttrue
\mciteSetBstMidEndSepPunct{\mcitedefaultmidpunct}
{\mcitedefaultendpunct}{\mcitedefaultseppunct}\relax
\EndOfBibitem
\bibitem{Lappi:2007ku}
T.~Lappi \href{http://dx.doi.org/10.1140/epjc/s10052-008-0588-4}{{\em Eur.
  Phys. J.} {\bf C55} (2008) 285} [\href{http://arXiv.org/abs/0711.3039}{{\tt
  arXiv:0711.3039 [hep-ph]}}]\relax
\mciteBstWouldAddEndPuncttrue
\mciteSetBstMidEndSepPunct{\mcitedefaultmidpunct}
{\mcitedefaultendpunct}{\mcitedefaultseppunct}\relax
\EndOfBibitem
\bibitem{Gardi:2006rp}
E.~Gardi, J.~Kuokkanen, K.~Rummukainen and H.~Weigert
  \href{http://dx.doi.org/10.1016/j.nuclphysa.2006.12.004}{{\em Nucl. Phys.}
  {\bf A784} (2007) 282} [\href{http://arXiv.org/abs/hep-ph/0609087}{{\tt
  arXiv:hep-ph/0609087 [hep-ph]}}]\relax
\mciteBstWouldAddEndPuncttrue
\mciteSetBstMidEndSepPunct{\mcitedefaultmidpunct}
{\mcitedefaultendpunct}{\mcitedefaultseppunct}\relax
\EndOfBibitem
\bibitem{Albacete:2007sm}
J.~L. Albacete \href{http://dx.doi.org/10.1103/PhysRevLett.99.262301}{{\em
  Phys. Rev. Lett.} {\bf 99} (2007) 262301}
  [\href{http://arXiv.org/abs/0707.2545}{{\tt arXiv:0707.2545 [hep-ph]}}]\relax
\mciteBstWouldAddEndPuncttrue
\mciteSetBstMidEndSepPunct{\mcitedefaultmidpunct}
{\mcitedefaultendpunct}{\mcitedefaultseppunct}\relax
\EndOfBibitem
\bibitem{Golec-Biernat:1998js}
K.~J. Golec-Biernat and M.~Wusthoff
  \href{http://dx.doi.org/10.1103/PhysRevD.59.014017}{{\em Phys. Rev.} {\bf
  D59} (1999) 014017} [\href{http://arXiv.org/abs/hep-ph/9807513}{{\tt
  arXiv:hep-ph/9807513}}]\relax
\mciteBstWouldAddEndPuncttrue
\mciteSetBstMidEndSepPunct{\mcitedefaultmidpunct}
{\mcitedefaultendpunct}{\mcitedefaultseppunct}\relax
\EndOfBibitem
\bibitem{Iancu:2003ge}
E.~Iancu, K.~Itakura and S.~Munier
  \href{http://dx.doi.org/10.1016/j.physletb.2004.02.040}{{\em Phys. Lett.}
  {\bf B590} (2004) 199} [\href{http://arXiv.org/abs/hep-ph/0310338}{{\tt
  arXiv:hep-ph/0310338}}]\relax
\mciteBstWouldAddEndPuncttrue
\mciteSetBstMidEndSepPunct{\mcitedefaultmidpunct}
{\mcitedefaultendpunct}{\mcitedefaultseppunct}\relax
\EndOfBibitem
\bibitem{Soyez:2007kg}
G.~Soyez \href{http://dx.doi.org/10.1016/j.physletb.2007.07.076}{{\em Phys.
  Lett.} {\bf B655} (2007) 32} [\href{http://arXiv.org/abs/0705.3672}{{\tt
  arXiv:0705.3672 [hep-ph]}}]\relax
\mciteBstWouldAddEndPuncttrue
\mciteSetBstMidEndSepPunct{\mcitedefaultmidpunct}
{\mcitedefaultendpunct}{\mcitedefaultseppunct}\relax
\EndOfBibitem
\bibitem{kuokkanen}
J.~Kuokkanen.
\newblock PhD thesis, University of Oulu, 2011\relax
\mciteBstWouldAddEndPuncttrue
\mciteSetBstMidEndSepPunct{\mcitedefaultmidpunct}
{\mcitedefaultendpunct}{\mcitedefaultseppunct}\relax
\EndOfBibitem
\end{mcitethebibliography}\endgroup
\bibliographystyle{JHEP-2modM}

\end{document}